\documentclass[reqno,11pt]{amsart}
\usepackage[linktocpage]{hyperref}
\usepackage[utf8]{inputenc}
\usepackage{graphicx}
\usepackage{slashed}
\usepackage{amscd}
\usepackage{amssymb}
\usepackage{esint}
\usepackage{enumitem}
\usepackage{braket}
\usepackage{comment}
\usepackage{amsmath}
\usepackage{dsfont}
\usepackage[mathscr]{eucal}
\usepackage{pstricks}
\usepackage{pst-all}

\numberwithin{equation}{section}
\allowdisplaybreaks[1]
\definecolor{labelkey}{gray}{.65}

\title[Notions of Fermionic Entropies for Causal Fermion Systems]{Notions of Fermionic Entropies \\ for
Causal Fermion Systems}

\author[F.\ Finster]{Felix Finster}
\address{Fakult\"at f\"ur Mathematik  Universit\"at Regensburg  D-93040 Regensburg  Germany}
\email{finster@ur.de}

\author[R.H. Jonsson]{Robert H. Jonsson}
\address{Nordita, KTH Royal Institute of Technology and Stockholm University, Hannes Alfv\'ens v\"ag 12, SE-106 91 Stockholm, Sweden}
\email{robert.jonsson@su.se}

\author[M.\ Lottner]{Magdalena Lottner}
\address{Fakult\"at f\"ur Mathematik  Universit\"at Regensburg  D-93040 Regensburg, Germany}
\email{magdalena.lottner@freenet.de}

\author[A.\ Much]{\\Albert Much}
\address{Institut f\"ur Theoretische Physik\\ Universit\"at Leipzig\\ D-04103 Leipzig, Germany}
\email{much@itp.uni-leipzig.de}

\author[S.\ Murro]{Simone Murro \\ \\ August 2024 / January 2025}
\address{Dipartimento di Matematica, Universit\`a di Genova and INFN Sezione di Genova, I-16146 Genova,  Italy}
\email{simone.murro@unige.it}

\newtheorem{Def}{Definition}[section]
\newtheorem{Thm}[Def]{Theorem}

\newtheorem{Lemma}[Def]{Lemma}

\newcommand{\Thanks}{\vspace*{.5em} \noindent \thanks}
\newcommand{\beq}{\begin{equation}}
\newcommand{\eeq}{\end{equation}}
\newcommand{\Proof}{\begin{proof}}
	\newcommand{\QED}{\end{proof} \noindent}

\newcommand{\la}{\langle}
\newcommand{\ra}{\rangle}

\newcommand{\Sl}{\mbox{$\prec \!\!$ \nolinebreak}}
\newcommand{\Sr}{\mbox{\nolinebreak $\succ$}}

\newcommand{\C}{\mathbb{C}}
\newcommand{\R}{\mathbb{R}}
\newcommand{\1}{\mathds{1}}
\newcommand{\Z}{\mathbb{Z}}
\newcommand{\N}{\mathbb{N}}

\DeclareMathOperator{\tr}{tr}

\renewcommand{\L}{{\mathcal{L}}}
\newcommand{\Sact}{{\mathcal{S}}}

\newcommand{\Cisc}{C^\infty_{\text{\rm{sc}}}}

\newcommand{\Dir}{{\mathcal{D}}}
\renewcommand{\H}{\mathscr{H}}

\newcommand{\Lin}{\text{\rm{L}}}
\newcommand{\F}{{\mathscr{F}}}

\DeclareMathOperator{\supp}{supp}

\newcommand{\scrM}{\mycal M}
\newcommand{\scrD}{\mycal D}
\newcommand{\scrN}{\mycal N}
\newcommand{\itemD}{\item[{\raisebox{0.125em}{\tiny $\blacktriangleright$}}]}

\newcommand{\rel}{\text{\rm{rel}}}

\newcommand{\s}{\mathfrak{s}}

\newcommand{\bitem}{\begin{itemize}[leftmargin=2.5em]}
\newcommand{\eitem}{\end{itemize}}

\newcommand{\id}{\text{id}}

\newcommand{\Fock}{{\mathcal{F}}}

\newcommand{\ii}{\mathrm{i}} 
\newcommand{\ee}[1]{\mathrm{e}^{#1}} 

\newcommand{\fermi}{{\mathrm{\tiny{f}}}}

\DeclareFontFamily{OT1}{rsfso}{}
\DeclareFontShape{OT1}{rsfso}{m}{n}{ <-7> rsfso5 <7-10> rsfso7 <10-> rsfso10}{}
\DeclareMathAlphabet{\mycal}{OT1}{rsfso}{m}{n}

\begin{document}

\begin{abstract}
The fermionic von Neumann entropy, the fermionic entanglement entropy and the fermionic relative entropy are defined for causal fermion systems. Our definition makes use of entropy formulas for quasi-free fermionic states in terms of the reduced one-particle density operator. Our definitions are illustrated in various examples for Dirac spinors in two- and four-dimensional Minkowski space, in the Schwarzschild black hole geometry and for fermionic lattices. We review area laws for the two-dimensional diamond and a three-dimensional spatial region in Minkowski space. The connection is made to the computation of the relative entropy using modular theory.
\end{abstract}

\maketitle

\vspace*{-0.5cm}

\tableofcontents

\section{Introduction} \label{secintro}
The purpose of this article is to study notions of fermionic entropy from the perspective of
causal fermion systems. As we shall see, the notions of fermionic von Neumann entropy,
entanglement entropy and relative entropy can be formulated naturally in this general setting.
Moreover, as we shall explain in various examples,
these notions of entropy reproduce the standard notions in various common settings.
This shows that causal fermion systems provide a general framework for studying
fermionic entropies, independent of the specific physical model in mind.
The main goal of this expository article is to explain how the common, well-known approaches to fermionic entanglement entropies can be formulated in and are related to the setting of causal fermion systems.

Entropy is a measure for the disorder of a physical system.
There are various notions of entropy, like the entropy in classical statistical mechanics
as introduced by Boltzmann and Gibbs, the Shannon and R{\'e}nyi entropies in information theory
or the {\em{von Neumann entropy}} for quantum systems.
In the past decade, there has been increasing interest in the {\em{entanglement entropy}}, 
which tells about quantum correlations between subsystems of a
composite quantum system~\cite{amico-fazio, horodecki}.
In the relativistic setting, the connection between modular theory
and the {\em{relative entropy}} has gained much attention
(see for example~\cite{hollands-sanders, galanda2023relative, witten-entangle}).

The theory of {\em{causal fermion systems}} is a recent approach to fundamental physics
(see the basics in Section~\ref{secprelim}, the reviews~\cite{dice2014, nrstg, review}, the textbooks~\cite{cfs, intro}
or the website~\cite{cfsweblink}).
In this approach, spacetime and all objects therein are described by a measure~$\rho$
on a set~$\F$ of linear operators on a Hilbert space~$(\H, \la .|. \ra_\H)$. 
The physical equations are formulated by means of the so-called {\em{causal action principle}}.
This is a nonlinear variational principle where an action~$\Sact$ is minimized under variations of the
measure~$\rho$. In different limiting cases, causal fermion systems give rise to the standard model of
particle physics and gravity on the level of classical field theory~\cite{cfs} and to quantum field
theory~\cite{fockfermionic, fockentangle, fockdynamics}.
Moreover, a general notion of quantum entropy was introduced for causal fermion systems in~\cite{entropy}.
In the present paper we focus on {\em{fermionic}} entropies.
Here we make use of the specific feature of causal fermion systems that, even in the fully interacting situation,
the physical system can be described by a family of fermionic one-particle wave functions.
This makes it possible to adapt notions for quasi-free fermionic states to causal fermion systems.

In order to explain this connection in more detail, we begin in the common description of a
quantum state by a {\em{density operator}}~$W$
on a fermionic Fock space~$\Fock$ (the density operator is often referred to as the {\em{statistical operator}};
here we do not use the standard notation~$\rho$ in order to avoid confusion with the measure
of the causal fermion system). A {\em{quasi-free and particle number preserving}}
fermionic state is fully characterized by its two-point distribution
\[ \omega_2(\overline{\psi}, \phi) := \tr_\Fock\big(\Psi^\dagger(\phi)\, \Psi(\overline{\psi})\, W \big) \]
(where~$\psi$, $\phi$ are one-particle wave functions and~$\Psi$ and~$\Psi^\dagger$ are the fermionic creation
and annihilation operators; for more details see the preliminaries in Section~\ref{secentquasi}).
Representing the two-point distribution relative to the scalar product on the one-particle Hilbert
space~$(\H, \la .|. \ra)$ gives the
{\em{reduced one-particle density operator}}~$D$, characterized by the relation
\[ \omega_2(\overline{\psi}, \phi) = \langle \psi | D \phi\rangle \:. \]
In this setting, the fermionic von Neumann entropy can be expressed in terms of the reduced
one-particle density operator by
\begin{align}
S &:= -\tr_\Fock \big( W \log W \big) \label{S1} \\
&\,= -\tr_\H \big( D\, \log D + (1-D)\, \log (1-D) \big) \:. \label{S2}
\end{align}
As we shall see, the last equation can be adapted to the general setting of causal fermion system
to serve as the definition of the fermionic von Neumann entropy.
Before going on, we point out that this method applies even in situations when the causal fermion system
describes a fully interacting physical system which, in the language of quantum field theory, does
{\em{not}} correspond to a quasi-free fermionic state (more specifically, the quantum state as
constructed in~\cite{fockfermionic, fockentangle} will in general not and does not need to be quasi-free).
The basic reason for this remarkable fact is that, in the setting of causal fermion systems, spacetime and all structures
therein are encoded in the one-particle wave functions of the system. Moreover, the above scalar product
has a counterpart in the so-called {\em{commutator inner product}}, making it possible to introduce
a one-particle density operator (for details see the preliminaries in Section~\ref{secprelimcfs}
and Section~\ref{secentropy}). This means that, even in fully interacting situations or in
physical systems in quantum spacetimes, the objects in~\eqref{S2} are well-defined,
making it possible to use this identity as the definition of the fermionic von Neumann entropy.

The method just described for the fermionic von Neumann entropy works similarly for the
fermionic entanglement entropy (Section~\ref{secententropy}) and the fermionic
relative entropy (Section~\ref{secrelative}). This makes it possible to analyze
various examples in detail, ranging from systems in two- and four-dimensional Minkowski
space (Sections~\ref{secminktwo}, \ref{secmink} as well as Sections~\ref{secrelmink} and~\ref{secreldiamond}),
the event horizon of a Schwarzschild black hole (Section~\ref{secbh}) to
fermionic lattice systems (Section~\ref{secspin}). The unifying theme is that all these examples
can be formulated in terms of causal fermion systems.
This shows that causal fermion systems provide a universal setting for the
formulation of fermionic entropies.

The paper is organized as follows. After providing the necessary preliminaries on
the fermionic entropies, causal fermion systems and briefly explaining the connection to modular theory (Section~\ref{secprelim}), the fermionic von Neumann entropy (Section~\ref{secentropy}) and the
corresponding fermionic entanglement entropy (Section~\ref{secententropy}) are introduced for causal fermion systems. These notions of entropy and entanglement entropy are worked out and explained in various examples:
two-dimensional Minkowski space and a causal diamond therein (Section~\ref{secminktwo}),
four-dimensional Minkowski space and a bounded spatial subset of a Cauchy surface therein (Section~\ref{secmink}),
the event horizon of the Schwarzschild black hole (Section~\ref{secbh})
and fermionic lattice systems (Section~\ref{secspin}), making a connection to condensed matter physics.
We proceed by introducing a corresponding notion of relative entropy (Section~\ref{secrelative}),
which we illustrate again in two-dimensional Minkowski space (Section~\ref{secrelmink})
and a causal diamond therein (Section~\ref{secreldiamond}).
The appendices provide the more technical background material.
Appendix~\ref{appA} is devoted to detailed proofs of how fermionic entropies 
can be expressed in terms of the reduced one-particle density operator. \\

\section{Preliminaries} \label{secprelim}
This section provides the necessary background on the entanglement entropy and on causal fermion systems.
Moreover, we explain the connection to modular theory.
\subsection{The Entanglement Entropy of a Quasi-Free Fermionic State} \label{secentquasi}
Given a Hilbert space~$(\H, \langle .|. \rangle$) (the ``one-particle Hilbert space''),
we let~$(\Fock, \la .|. \ra_\Fock)$ be the corresponding fermionic Fock space, i.e.,
\[ \Fock = \bigoplus_{k=0}^N \;\underbrace{\H \wedge \cdots \wedge \H}_{\text{$k$ factors}} \]
(where~$\wedge$ denotes the totally anti-symmetrized tensor product).
We define the {\em{creation operator}}~$\Psi^\dagger$ by
\[ \Psi^\dagger \::\: \H \rightarrow \Lin(\Fock) \:,\qquad
\Psi^\dagger(\psi) \big( \psi_1 \wedge \cdots \wedge \psi_p \big) := \psi \wedge \psi_1 \wedge \cdots \wedge \psi_p \]
(where~$\Lin(\Fock)$ denotes the bounded linear operators on~$\Fock$).
The adjoint of~$\Psi^\dagger(\psi)$ is the annihilation operator denoted by~$\Psi(\overline{\psi}) := (\Psi^\dagger(\psi))^*$. These operators satisfy the canonical anti-commutation relations
\[ 
	\big\{ \Psi(\overline{\psi}), \Psi^\dagger(\phi) \big\} = \langle \psi | \phi \rangle \qquad\text{and}\qquad
	\big\{ \Psi(\overline{\psi}), \Psi(\overline{\phi}) \big\} = 0 = \big\{ \Psi^\dagger(\psi), \Psi^\dagger(\phi) \big\} \:. \]
Next, we let~$W$ be a {\em{density operator}} (or {\em{statistical operator}}) on~$\Fock$, i.e., a positive
semi-definite linear operator of trace one,
\[ W \::\: \Fock \rightarrow \Fock\:,\qquad W \geq 0 \quad \text{and} \quad \tr_\Fock(W)=1 \:. \]
Given an observable~$A$ (i.e., a symmetric operator on~$\Fock$), the expectation value of the measurement
is given by
\[ \langle A \rangle := \tr_\Fock\big( A W) \:. \]
In the algebraic formulation, the corresponding {\em{quantum state}}~$\omega$ is defined as
the linear functional which to every observable associates its expectation value, i.e.,
\[ \omega \::\: A \mapsto \tr_\Fock\big( A W) \:. \]

In this paper, we restrict our attention to the subclass of so-called \emph{quasi-free}
states, also referred to as {\em{Gaussian states}}, which are fully determined by their two-point distributions.
More precisely, for a quasi-free state all odd $n$-point distributions vanish, whereas all even $n$-point
distributions can be computed using Wick's theorem.
Moreover, we restrict attention to the state is {\em{particle-number preserving}}, meaning that
all two-point expectations involving two creation or two annihilation operators vanish,
\[  \omega\big( \Psi^\dagger(\phi)\, \Psi^\dagger(\psi) \big) = 0 =  \omega\big( \Psi(\overline{\phi})\, \Psi(\overline{\psi}) \big) \:. \]
In the literature, this property is sometimes referred to as a {\em{gauge-invariant state}}
(see~\cite[Proposition~17.32]{derzinski-gerard-quantum}).
A quasi-free and particle-number preserving state is fully determined by its two-point distributions
\beq \label{omega2def}
\omega_2(\overline{\psi}, \phi) := \omega\big( \Psi^\dagger(\phi)\, \Psi(\overline{\psi}) \big) \:.
\eeq

\begin{Def} \label{defreduced}
The {\bf{reduced one-particle density operator}}~$D$ is the positive semi-definite linear operator on the Hilbert space~$(\H, \la .|. \ra)$ defined by
\beq \label{Ddef}
\omega_2(\overline{\psi}, \phi) = \langle \psi | D \phi\rangle \qquad \text{for all~$\psi, \phi \in \H$} \:.
\eeq
\end{Def}

The {\em{von Neumann entropy}}~$S$ of the quasi-free and particle number preserving fermionic
state~$\omega$ can be expressed in terms of the reduced one-particle density operator by
\beq \label{Sred}
S(\omega) = \tr \eta(D) \:,
\eeq
where~$\eta$ is the \textit{von Neumann entropy function} defined by
\begin{flalign}
\label{etadef}
\eta(x):= \begin{cases}
	-x\log x - (1-x) \log (1-x)& \quad \text{if~$x \in (0,1)$} \\
	0 & \quad \text{otherwise}\:.
\end{cases}
\end{flalign}
This formula appears commonly in the literature
(see for example~\cite[Equation 6.3]{ohya-petz}, \cite{klich, casini-huerta, longo-xu}
and~\cite[eq.~(34)]{helling-leschke-spitzer}).
A detailed derivation is given in Appendix~\ref{appA} (see Theorem~\ref{thmreduced}).

For the {\em{entanglement entropy}} we need to consider a subsystem of our quantum system.
For our purpose, it is sufficient to consider systems formed of wave functions in 
space~$\scrN$ (which could be one-dimensional space~$\R$, three-dimensional space~$\R^3$,
a three-dimensional manifold or a lattice).
Given a spatial subregion~$\Lambda \subset \scrN$, we consider the entropic difference
\beq \label{entropygen}
S_\Lambda(D) := \tr \Big( 
\eta \big( \chi_{\Lambda} \:D\: \chi_{\Lambda} \big) -\chi_{\Lambda} \,\eta(D)\,\chi_{\Lambda}
\Big) \:,
\eeq
where~$\chi_\Lambda$ is the operator of multiplication by the characteristic function.
Also this formula appears commonly in the literature. 
A detailed derivation is given in Appendix~\ref{appA} (see Theorem~\ref{thmentanglement}).
Following the conventions in the mathematical physics literature, in what follows we refer to~$S_\Lambda$ as the
{\em{fermionic entanglement entropy}}. More details can be found in~\cite[Section~3]{leschke-sobolev-spitzer2}.

\subsection{Causal Fermion Systems and the Causal Action Principle} \label{secprelimcfs}
We now recall a few basics on causal fermion systems, restricting attention to the structures
needed in the sequel.
\subsubsection{Basic Definitions}
We begin with the general definitions.
\begin{Def} \label{defcfs} (causal fermion systems) {\em{ 
Given a separable complex Hilbert space~$\H$ with scalar product~$\la .|. \ra_\H$
and a parameter~$n \in \N$ (the {\em{spin dimension}}), we let~$\F \subset \Lin(\H)$ be the set of all
symmetric operators on~$\H$ of finite rank, which (counting multiplicities) have
at most~$n$ positive and at most~$n$ negative eigenvalues. On~$\F$ we are given
a positive measure~$\rho$ (defined on a $\sigma$-algebra of subsets of~$\F$).
We refer to~$(\H, \F, \rho)$ as a {\em{causal fermion system}}.
}}
\end{Def} \noindent
A causal fermion system describes a spacetime together
with all structures and objects therein.
The physical equations are formulated for a causal fermion system
by demanding that the measure~$\rho$ should be a minimizer of the causal action principle,
which we now introduce. For brevity of the presentation, we only consider the
{\em{reduced causal action principle}} where the so-called boundedness constraint has been
incorporated by a Lagrange multiplier term. This simplification is no loss of generality, because
the resulting EL equations are the same as for the non-reduced action principle
as introduced for example~\cite[Section~\S1.1.1]{cfs}.

For any~$x, y \in \F$, the product~$x y$ is an operator of rank at most~$2n$. 
However, in general it is no longer a symmetric operator because~$(xy)^* = yx$,
and this is different from~$xy$ unless~$x$ and~$y$ commute.
As a consequence, the eigenvalues of the operator~$xy$ are in general complex.
We denote the rank of~$xy$ by~$k \leq 2n$. Counting algebraic multiplicities, we choose~$\lambda^{xy}_1, \ldots, \lambda^{xy}_{k} \in \C$ as all the non-zero eigenvalues and set~$\lambda^{xy}_{k+1}, \ldots, \lambda^{xy}_{2n}=0$.
Given a parameter~$\kappa>0$ (which will be kept fixed throughout this paper),
we introduce the $\kappa$-Lagrangian and the causal action by
\begin{align}
\text{\em{$\kappa$-Lagrangian:}} && \L(x,y) &= 
\frac{1}{4n} \sum_{i,j=1}^{2n} \Big( \big|\lambda^{xy}_i \big|
- \big|\lambda^{xy}_j \big| \Big)^2 + \kappa\: \bigg( \sum_{j=1}^{2n} \big|\lambda^{xy}_j \big| \bigg)^2 \label{Lagrange} \\
\text{\em{causal action:}} && \Sact(\rho) &= \iint_{\F \times \F} \L(x,y)\: d\rho(x)\, d\rho(y) \:. \label{Sdef}
\end{align}
The {\em{reduced causal action principle}} is to minimize~$\Sact$ by varying the measure~$\rho$
under the following constraints,
\begin{align}
\text{\em{volume constraint:}} && \rho(\F) = 1 \quad\;\; \label{volconstraint} \\
\text{\em{trace constraint:}} && \int_\F \tr(x)\: d\rho(x) = 1 \:. \label{trconstraint}
\end{align}
This variational principle is mathematically well-posed if~$\H$ is finite-dimensional.
For a review of the existence theory and the analysis of general properties of minimizing measures
we refer to~\cite[Chapter~12]{intro}.

A minimizer of the causal action principle
satisfies the following {\em{Euler-Lagrange (EL) equations}}.
For a suitable value of the parameter~$\s>0$,
the function~$\ell : \F \rightarrow \R_0^+$ defined by
\beq \label{elldef}
\ell(x) := \int_M \L(x,y)\: d\rho(y) - \s
\eeq
is minimal and vanishes on the support of~$\rho$,
\[ 
\ell|_{\supp \rho} \equiv \inf_\F \ell = 0 \:. \]
Likewise, the parameter~$\s \geq 0$ in~\eqref{elldef} is the Lagrange parameter
corresponding to the volume constraint. For the derivation and further details we refer to~\cite[Section~2]{jet}
or~\cite[Chapter~7]{intro}.

\subsubsection{Spacetime and Causal Structure}
Let~$\rho$ be a {\em{minimizing}} measure. {\em{Spacetime}}
is defined as the support of this measure,
\beq \label{Mdef}
M := \supp \rho \;\subset\; \F \:,
\eeq
where on~$M$ we consider the topology induced by~$\F$ (generated by the operator norm
on~$\Lin(\H)$).
Thus the spacetime points are symmetric linear operators on~$\H$.
The restriction of the measure~$\rho|_M$ gives a volume measure on spacetime.

The operators in~$M$ contain a lot of information which, if interpreted correctly,
gives rise to spacetime structures like causal and metric structures, spinors
and interacting fields (for details see~\cite[Chapter~1]{cfs}).
All the resulting objects are {\em{inherent}} in the sense that we only use information
already encoded in the causal fermion system.
Here we restrict attention to those structures needed in what follows.
We begin with the following notion of causality:

\begin{Def} (causal structure) \label{def2} 
{\em{ For any~$x, y \in \F$, we again denote the non-trivial ei\-gen\-values of the operator product~$xy$
(again counting algebraic multiplicities) by~$\lambda^{xy}_1, \ldots, \lambda^{xy}_{2n}$.
The points~$x$ and~$y$ are
called {\em{spacelike}} separated if all the~$\lambda^{xy}_j$ have the same absolute value.
They are said to be {\em{timelike}} separated if the~$\lambda^{xy}_j$ are all real and do not all 
have the same absolute value.
In all other cases (i.e., if the~$\lambda^{xy}_j$ are not all real and do not all 
have the same absolute value),
the points~$x$ and~$y$ are said to be {\em{lightlike}} separated. }}
\end{Def} \noindent
Restricting the causal structure of~$\F$ to~$M$, we get causal relations in spacetime.

The Lagrangian~\eqref{Lagrange} is compatible with the above notion of causality in the
following sense.
Suppose that two points~$x, y \in M$ are spacelike separated.
Then the eigenvalues~$\lambda^{xy}_i$ all have the same absolute value.
As a consequence, the Lagrangian~\eqref{Lagrange} vanishes. Thus pairs of points with spacelike
separation do not enter the action. This can be seen in analogy to the usual notion of causality where
points with spacelike separation cannot influence each other.
This is the reason for the notion ``causal'' in {\em{causal}} fermion system
and {\em{causal}} action principle.

\subsubsection{Spinors and Physical Wave Functions} \label{secinherent}
A causal fermion system also gives rise to spinorial wave functions in spacetime, as we now explain.
For every~$x \in \F$ we define the {\em{spin space}}~$S_x$ by~$S_x = x(\H)$;
it is a subspace of~$\H$ of dimension at most~$2n$.
It is endowed with the {\em{spin inner product}}~$\Sl .|. \Sr_x$ defined by
\[ 
\Sl u | v \Sr_x = -\la u | x v \ra_\H \qquad \text{(for all~$u,v \in S_x$)} \:. \]
It is an important observation that every vector~$u \in \H$ of the Hilbert space gives rise to a unique
wave function denoted by~$\psi^u$, which to every~$x \in M$ associates a vector of the
corresponding spin space~$\psi^u(x) \in S_x$. It is obtained by orthogonal projection to the
spin space,
\[ 
\psi^u \::\: M \rightarrow \H\qquad \text{with} \qquad \psi^u(x) := \pi_x u \in S_xM \quad \text{for all~$x \in M$}\:. \]
We refer to~$\psi^u$ as the {\em{physical wave function}} of the vector~$u \in \H$.
Varying the vector~$u \in \H$, we obtain a whole family of physical wave functions.
This family is described most conveniently by the
{\em{wave evaluation operator}}~$\Psi$ defined at every spacetime point~$x \in M$ by
\[ 
\Psi(x) \::\: \H \rightarrow S_x \:, \qquad u \mapsto \psi^u(x) \:. \]
It is a simple but important observation that every spacetime point operator can be recovered from
its wave evaluation operator by (for the proof see for example~\cite[Lemma~1.1.3]{cfs}).
\[ 
x = - \Psi(x)^* \Psi(x) \:. \]
Having constructed the spacetime point operators, we also recover all the other
inherent structures of a causal fermion system. Proceeding in this way, all spacetime structures
can be regarded as being induced by the physical wave functions.
Moreover,  restricting attention to variations of~$\Psi$, one can understand the causal action principle as
a variational principle for the family of physical wave functions.
Finally, one can construct concrete examples
of causal fermion systems by choosing the physical wave functions more specifically as
the quantum mechanical wave functions
in a classical Lorentzian spacetime. In the next section we explain this construction in more detail.

\subsubsection{Surface Layer Integrals and the Commutator Inner Product} \label{secHextend}
In the setting of causal fermion systems, integrals over hypersurfaces
are replaced by so-called {\em{surface layer integrals}}, which are
double integrals of the general form
\beq \label{osi}
\int_\Omega \bigg( \int_{M \setminus \Omega} (\cdots)\: \L(x,y)\: d\rho(y) \bigg)\, d\rho(x) \:,
\eeq
where~$(\cdots)$ stands for suitable variational derivatives of the Lagrangian, and~$\Omega$
is a Borel subset of~$M$.
The connection can be understood most easily in the case when~$\L(x,y)$ vanishes
unless~$x$ and~$y$ are close together. In this case, we only get a contribution to~\eqref{osi}
if both~$x$ and~$y$ are close to the boundary of~$\Omega$.
A more detailed explanation of the idea of a surface layer integrals is given in~\cite[Section~2.3]{noether}.

Surface layer integrals were first introduced in~\cite{noether} in order to 
make a connection between symmetries and conservation laws for surface layer integrals.
Here we will make essential use of the conservation law corresponding to the
symmetry under unitary transformations on the Hilbert space~$\H$.
For a minimizing measure~$\rho$, it gives rise to a conservation law for a sesquilinear form on the physical wave functions of the form
\beq \label{OSIdyn}
\begin{split}
\la u | v \ra^t_\rho = -2i \,\bigg( \int_{\Omega^t} \!d\rho(x) \int_{M \setminus \Omega^t} \!\!\!\!\!\!\!d\rho(y) 
&- \int_{M \setminus \Omega^t} \!\!\!\!\!\!\!d\rho(x) \int_{\Omega^t} \!d\rho(y) \bigg)\\
&\times\:
\Sl \psi^u(x) \:|\: Q(x,y)\, \psi^v(y) \Sr_x \:,
\end{split}
\eeq
where~$\Omega_t$ can be thought of as the past of a Cauchy surface.
Here ``conservation law'' means that this inner product is independent of~$t$.
The sesquilinear form~\eqref{OSIdyn} is referred to as the {\em{commutator inner product}}
(the name comes from the fact that the unitary invariance can be expressed in terms of
commutators; see~\cite[Section~3]{dirac} for details).
The kernel~$Q(x,y)$ appearing in this formula is the first variational derivative of the Lagrangian
 (see~\cite{dirac} for details).
 
 In~\cite[Section~5]{noether} it was shown that, taking the continuum limit of the vacuum in Minkowski space,
this sesquilinear form coincides, up to a constant, with the scalar product~$\la u|v\ra_\H$.
We now give this property a useful name. We only assume that this property holds for all vectors in
a finite-dimensional subspace~$\H^\fermi \subset \H$ (for details see~\cite[Appendix~A]{current}).
The vectors in~$\H^\fermi$ can be regarded of as the low-frequency wave functions, i.e.\ those wave functions
whose energy and momentum is very small on the Planck scale. Alternatively, one can think of~$\H^\fermi$
as being composed all wave functions which are accessible to experiments.

\begin{Def} \label{defSLrep}
Given a critical measure~$\rho$ and a past set~$\Omega_t \subset M$, the commutator inner product
is said to {\bf{represent the scalar product on}} the finite-dimensional subspace~{\bf{$\H^\fermi \subset \H$}} if
\beq \label{Ccond}
\la u|v \ra^t_\rho = c\, \la u|v \ra_\H \qquad \text{for all~$u,v \in \H^\fermi$}
\eeq
with a suitable positive constant~$c$.
\end{Def}

\subsection{Connection to Modular Theory}
The fermionic entanglement entropy as studied in the present paper is closely related to the
modular theory. We now explain this connection, beginning with a short concise introduction
to the basics of modular theory (for more details see~\cite{borchers, haag1996local, witten-entangle}).
Let~$\mathcal{H}$ be a Hilbert space and~$\mathcal{M}$ be a von Neumann algebra acting on this space. We denote by~$\mathcal{M}'$  the commutant of~$\mathcal{M}$. Furthermore, we call a vector~$\Omega$ cyclic and separating if~$\mathcal{M}'\Omega$ are dense in~$\mathcal{H}$. If these conditions are fulfilled, then there exist two anti-linear operators (called the Tomita operators)~$S:\mathcal{H}\supseteq\text{dom}(S)\rightarrow\mathcal{H}$ and its adjoint~$S^*:\mathcal{H}\supseteq\text{dom}(S^*)\rightarrow\mathcal{H}$ that have the property
\begin{align*}
     SA\,\Omega&= A^*\Omega \qquad \qquad \forall A\in\mathcal{M}\\
     S^*B\,\Omega&= B^*\Omega \qquad \qquad \forall B\in\mathcal{M}' \:.
\end{align*}
Since~$S^2=1$ the operator is invertible and thus has a {\em{unique}} polar decomposition 
\[ S=J\Delta^{\frac{1}{2}} \:, \]
where~$\Delta:\mathcal{H}\supseteq\text{dom}(\Delta)\rightarrow\mathcal{H}$ is denoted as the modular operator and the partial isometry~$J:\mathcal{H}\rightarrow\mathcal{H}$ is called the modular conjugation associated to the pair~$(\mathcal{M},\Omega)$. The modular conjugation   maps the von Neumann algebra to its commutant, i.e.\ 
\[ J\mathcal{M}J=\mathcal{M}' \:, \]
and  the modular operator (also called the modular Hamiltonian) is selfadjoint, positive and invertible and defines a group of automorphisms of the von Neumann algebra~$\mathcal{M}$ and the commutant thereof,
\[ \Delta^{it}\mathcal{M}^{(')}\Delta^{-it}=\mathcal{M}^{(')} \:, \]
for all~$t\in\R$. This group is used to formulate the KMS-condition and to define the Araki-Uhlmann relative entropy. For the case where the state (or vector representative)~$\Psi$ and~$\Omega$ are unitarily related, let us say by the operator~$U$ (where the domain of~$\Delta^{\frac{1}{2}}$ is stable under the application of the unitary operator, $U\text{dom}(\Delta^{\frac{1}{2}})\subset \Delta^{\frac{1}{2}}$) the Araki-Uhlmann relative entropy takes the form 
\[ S(\Psi,\Omega)=-\langle \Psi, \log(\Delta)\,\Psi\rangle \:. \]
This quantity was calculated in various instances in quantum field theory \cite{longo, longo-xu, hollands-entropy, hollands-ishibashi, dangelo,  kurpicz, galanda2023relative, froeb, froeb-much, dangelo-froeb, cadamuro-froeb},
also in relation with the Bekenstein bound~\cite{casini}. 

The connection to the fermionic entanglement entropy is obtained by the fundamental relation between
the modular operator~$\Delta$ and the one-particle density operator~$D$ as introduced in
Definition~\ref{defreduced} given by
\beq \label{DelD}
D = (1+\Delta)^{-1} \:,
\eeq
where~$\Delta=\exp(-\mathcal{H})$, with~$\mathcal{H}$ being the modular Hamiltonian, c.f.~\cite{araki1970quasifree}, \cite[Section 3.3]{longo-xu} and \cite{casini-huerta2}
(in \cite{longo-xu}, the operator~$D$ is referred to as the covariance operator~$C$).
In the example of the Rindler wedge, the modular Hamiltonian is given by~$2\pi\,K$, where~$K$ is the 
generator of Lorentz boosts. This result is known as the Bisognano-Wichmann
theorem~\cite{bisognano-wichmann1, bisognano-wichmann2}.

The simple formula~\eqref{DelD} makes it possible to relate all our results on the
one-particle density operator~$D$ to corresponding results for the modular operator.
Moreover, our methods and results for fermionic entropies complement the techniques
from modular theory. We hope that exploring these connections further will be fruitful for the future
development of the theories.

\section{The Fermionic Von Neumann Entropy of a Causal Fermion System} \label{secentropy}
In what follows, we shall {\em{not}} assume that the commutator inner product represents the
scalar product (see Definition~\ref{defSLrep}). Instead, we merely assume that
this sesquilinear form is positive semi-definite and bounded with respect to the scalar product, i.e.,
\[ 0 \leq \la u|u \ra^t_\rho \leq c\: \|u\|_\H^2 \qquad \text{for all~$u \in \H^\fermi$} \]
with~$c$ as in~\eqref{Ccond}. Here, as explained before Definition~\ref{defSLrep}, the
subspace~$\H^\fermi \subset \H$ is formed of all low-frequency wave functions.
Using the Fr{\'e}chet-Riesz theorem, there is a unique operator~$\sigma \in \Lin(\H^\fermi)$ such that
\beq \label{sigmadef}
\la u|v \ra^t_\rho = \la u | \sigma v \ra_\H \qquad \text{for all~$u,v \in \H^\fermi$} \:.
\eeq
Clearly, the operator~$\sigma$ is positive semi-definite and bounded.
By a scaling of the Hilbert scalar product~$\la .|. \ra_\H$, we can arrange that its norm is at most one, i.e.,
\beq \label{sigmain}
0 \leq \sigma \leq 1 \:.
\eeq
In many situations it seems a good idea to arrange by scaling that its norm is even equal to one,
but this will not needed for the subsequent constructions.
Now we can take the identity~\eqref{Sdef} as the definition of the fermionic entropy.
\begin{Def} \label{defvonNeumann}
We define the {\bf{fermionic von Neumann entropy}}~$S$ of the causal fermion system~$(\H, \F, \rho)$ by
\[ S = \tr_{\H^\fermi} \big(\eta(\sigma) \big) \:, \]
where~$\eta$ is again the von Neumann entropy function~\eqref{etadef}.
\end{Def}
If the commutator inner product represents the scalar product, then~$\sigma=\1$, and
the fermionic entropy vanishes. As shown in~\cite[Section~5]{noether}, this is the case in the Minkowski vacuum.

\section{The Fermionic Entanglement Entropy of a Subsystem} \label{secententropy}
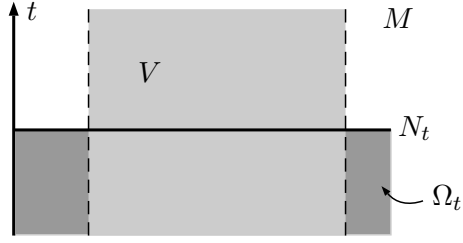
\begin{figure}
%
\psscalebox{1.0 1.0} 
{
\begin{pspicture}(-4.2,-1.5590234)(10.055,1.5590234)
\definecolor{colour0}{rgb}{0.8,0.8,0.8}
\definecolor{colour1}{rgb}{0.6,0.6,0.6}
\psframe[linecolor=colour0, linewidth=0.02, fillstyle=solid,fillcolor=colour1, dimen=outer](5.02,-0.14902344)(0.03,-1.5590234)
\psframe[linecolor=colour0, linewidth=0.02, fillstyle=solid,fillcolor=colour0, dimen=outer](4.41,1.4409766)(1.01,-1.5590234)
\psline[linecolor=black, linewidth=0.04, arrowsize=0.05291667cm 2.0,arrowlength=1.4,arrowinset=0.0]{->}(0.02,-1.5590234)(0.02,1.5409765)
\psline[linecolor=black, linewidth=0.04](0.01,-0.15902343)(5.01,-0.15902343)
\rput[bl](5.105,-0.29402342){\normalsize{$N_t$}}
\rput[bl](0.195,1.3009765){\normalsize{$t$}}
\rput[bl](4.9,1.1909766){\normalsize{$M$}}
\rput[bl](1.665,0.48097655){\normalsize{$V$}}
\psline[linecolor=black, linewidth=0.02, linestyle=dashed, dash=0.17638889cm 0.10583334cm](1.01,1.4409766)(1.01,-1.5590234)
\psline[linecolor=black, linewidth=0.02, linestyle=dashed, dash=0.17638889cm 0.10583334cm](4.41,1.4409766)(4.41,-1.5590234)
\psbezier[linecolor=black, linewidth=0.02, arrowsize=0.05291667cm 2.0,arrowlength=1.4,arrowinset=0.0]{->}(5.435,-1.0990235)(5.24,-1.1640234)(5.075,-1.1290234)(4.885,-0.9040234375)
\rput[bl](5.555,-1.2140235){\normalsize{$\Omega_t$}}
\end{pspicture}
}
\caption{Spatial localization of the scalar product.}
\label{figfoliation}
\end{figure}

We now choose an open subset~$V \subset M$ and ``localize'' in~$V \cap N_t$
by setting (see Figure~\ref{figfoliation})

\begin{align}
\la \psi | \phi \ra^t_{V,\rho} := -2i \bigg( \int_{\Omega^t \cap V} d\rho(x) \int_{M \setminus \Omega^t} d\rho(y)
&- \int_{\Omega^t} d\rho(x) \int_{V \setminus \Omega^t} d\rho(y) \bigg) \notag \\
&\times\: \Sl \psi(x) \:|\: Q(x,y)\, \phi(y) \Sr_x \:, \label{osiloc}
\end{align}
where~$N_t = \partial \Omega_t$ can be thought of as the Cauchy surface.
Next, we represent this localized scalar product similar to~\eqref{sigmadef} in terms of the Hilbert space scalar product,
\beq \label{sigmaVdef}
\la u|v \ra^t_{V,\rho} = \la u | \sigma_V v \ra_\H \qquad \text{for all~$u,v \in \H^\fermi$} \:.
\eeq
We assume that~$\sigma_V$ is again positive and bounded by~$\sigma$. Then,
in combination with~\eqref{sigmain} we have
\beq \label{sigmaVin}
0 \leq \sigma_V \leq \sigma \leq 1 \:.
\eeq
In the case~$\sigma=\1$ that the total system is described by a pure fermionic state, the
entanglement entropy of the subsystem is defined as its von Neumann entropy, i.e.\
$S_V = \tr_{\H^\fermi} \big(\eta(\sigma_V))$. In the more general case that the total system is in
a mixed fermionic state, similar to~\eqref{entropygen} one needs to subtract a corresponding volume
contribution. However, in the context of causal fermion systems, we cannot define the
entanglement entropy using the formula~\eqref{entropygen} by~$S_\Lambda(\sigma)$,
because multiplying by the characteristic function does in general not map back to the Hilbert
space~$\H$. In order to circumvent this problem, one introduces the operator~$\chi_V$ by
\beq \label{chiVdef}
\chi_V := \big( \sigma^{-\frac{1}{2}} \,\sigma_V\, \sigma^{-\frac{1}{2}} \big)^\frac{1}{2} \:,
\eeq
making it possible to define the entanglement entropy as~$S_\Lambda(\sigma)$.

The formula~\eqref{chiVdef} can be motivated and understood as follows. First of all, the operator on the right
is obviously symmetric, and its norm is bounded by one. Moreover,
by a direct computation starting from~\eqref{chiVdef} one sees that the operator~$\sigma_V$
satisfies the relation
\[ \sigma_V = \sigma^{\frac{1}{2}} \, \chi_V^2 \, \sigma^{\frac{1}{2}} \:. \]
Using that the spectrum of an operator product is invariant under cyclic commutation of the factors, one sees
that the operator~$\sigma_V$ is isospectral to the operator product~$\chi_V \sigma \chi_V$,
\[ \sigma_V \simeq \chi_V\, \sigma\, \chi_V \:. \]
Therefore, at least formally, $\tr_{\H^\fermi} \big(\eta(\sigma_V)) = 
\tr_{\H^\fermi} \big(\eta(\chi_V\, \sigma\, \chi_V))$, giving us the 
first term of the entropic difference~\eqref{entropygen}. This explains why our choice of~$\chi_V$
makes sense and also motivates how to choose the counter term.
This leads us to the following definition.

\begin{Def} \label{defentangle}
We introduce the {\bf{fermionic entanglement entropy}} by
\beq \label{SVdef}
S_{V} = \tr_{\H^\fermi} \big(\eta(\sigma_V) - \chi_V \,\eta(\sigma)\, \chi_V \big) \:,
\eeq
where~$\chi_V$ is defined by~\eqref{chiVdef}, and~$\eta$ is again the von
Neumann entropy function~\eqref{etadef}.
\end{Def} \noindent

\section{Example: Two-Dimensional Minkowski Space} \label{secminktwo}
In this section we consider the example of a causal fermion system describing the Minkowski vacuum
in two spacetime dimensions.

We recall a few basics, using the notation in~\cite{rindler}.
Let~$(\scrM, g)$ be two-dimensional Minkowski space, i.e.~$\scrM = \R^{1,1}$
with the line element
\[ 
ds^2 = g_{ij}\: dx^i dx^j = dt^2 - dx^2 \:. \]
Moreover, we let~$S \scrM = \scrM \times \C^2$
be the trivial spinor bundle, endowed with the {\em{spin inner product}} defined by
\beq \label{ssprod}
\Sl \psi | \phi \Sr = \la \psi, \begin{pmatrix} 0 & 1 \\ 1 & 0 \end{pmatrix} \phi \ra_{\C^2}
\eeq
(where~$\la .,. \ra_{\C^2}$ is the canonical scalar product on~$\C^2$).
We work in the so-called chiral representation of the Dirac matrices
\[ 
\gamma^0 = \begin{pmatrix} 0 & 1 \\ 1 & 0 \end{pmatrix} \:,\qquad
\gamma^1 = \begin{pmatrix} 0 & 1 \\ -1 & 0 \end{pmatrix} \:. \]
The Dirac matrices are symmetric with respect to the spin inner product~\eqref{ssprod}.
The spin scalar product is an indefinite inner product of signature~$(1,1)$.
Introducing the {\em{Dirac operator}}
\beq \label{Dirop}
\Dir := i \gamma^j \partial_j \:,
\eeq
the {\em{massive Dirac equation}} reads
\beq \label{Direq}
(\Dir-m) \psi = 0 \:,
\eeq
where~$m>0$ is the rest mass (we always work in natural units~$\hbar=c=1$).
Taking smooth and compactly supported initial data on a Cauchy surface~$\scrN$
and solving the Cauchy problem, one obtains a Dirac solution in the class~$\Cisc(\scrM, S\scrM)$
of smooth wave functions with spatially compact support. On solutions~$\psi, \phi$
in this class, one defines the (positive definite) scalar product
\[ 
(\psi | \phi)_m := \int_\scrN \Sl \psi | \slashed{\nu} \phi\Sr|_q\: d\mu_\scrN(q) \:, \]
where~$\slashed{\nu} = \gamma^j \nu_j$ denotes Clifford multiplication by the future-directed unit
normal~$\nu$, and~$d\mu_\scrN$ is the volume measure of the induced Riemannian metric
on~$\scrN$ (thus for the above ray~$\scrN = \{ (\alpha x, x) \text{ with } x > 0 \}$,
the measure~$d\mu_\scrN= \sqrt{1-\alpha^2}\, dx$ is a multiple of the Lebesgue measure).
Due to current conservation, this scalar product is independent of the choice of~$\scrN$.
Forming the completion, we obtain the Hilbert space~$(\H_m, (.|.)_m)$, referred to as the
{\em{solution space}} of the Dirac equation.
For convenience, we always
choose~$\scrN$ as the Cauchy surface~$\{t=0\}$, so that
\beq \label{printM}
(\psi | \phi)_m = \int_{-\infty}^\infty \Sl \psi | \gamma^0 \phi \Sr|_{(0,x)}\: dx \:.
\eeq

\subsection{Construction of the Causal Fermion System} \label{secmink1}
For the construction of the causal fermion system, we must choose a closed subspace~$\H \subset \H_m$,
having the interpretation as the occupied one-particle states of the system.
In order to describe the fermionic vacuum, one chooses~$\H$ as the subspace of all
negative-frequency solutions of the Dirac equation.
One way of introducing this subspace is to write the negative-frequency solutions as Fourier integrals,
\beq \label{Fourierneg}
\psi(t, x) = \int_{\R^2} \frac{d^2k}{(2 \pi)^2}\: (k_j \gamma^j + m)\: \delta\big( k^2+m^2\big) \: \Theta(-k_0)\:
\hat{\psi}\big(k_1 \big) \: e^{i k x}
\eeq
with~$\hat{\psi} \in C^\infty_0(\R, \C^2)$
(note that the Heaviside function~$\Theta(-k^0)$ has the effect that only negative frequencies are considered).
Taking the closure of all these wave functions gives the subspace~$\H \subset \H_m$.
For clarity, we denote the scalar product on this subspace by~$\la .|. \ra_\H := (.|.)_m|_{\H \times \H}$.

We point out that the functions in~$\H$ are in general not continuous.
Therefore, we cannot evaluate the wave functions pointwise at a spacetime point~$x \in \scrM$.
However, for the following constructions it is crucial to do so.
The way out is to introduce so-called {\em{regularization operators}}~$({\mathfrak{R}}_\varepsilon)_\varepsilon$
with~$0 < \varepsilon < \varepsilon_{\max}$
which map~$\H$ to the continuous wave functions,
\beq \label{Repsdef}
{\mathfrak{R}}_\varepsilon \::\: \H \rightarrow C^0(\scrM, S\scrM) \:.
\eeq
In the limit~$\varepsilon \searrow 0$, these operators should go over to the identity
(in a suitable sense which we do not need to specify here).
The physical picture is that on a small length scale, which can be thought of as the
Planck length scale~$\varepsilon \approx 10^{-35}$~meters, the structure of spacetime
must be modified. The regularization operators specify this microscopic
structure of spacetime. Many different choices of regularization operators are possible.
In order to keep the presentation as simple as possible, we here restrict attention to
the so-called~$i \varepsilon$-regularization, where we insert an exponentially decaying factor~$e^{\varepsilon k_0}$
into the Fourier integral~\eqref{Fourierneg}, i.e.
\beq \label{Fourierreg}
({\mathfrak{R}}_\varepsilon \psi)(t, \vec{x}) := \int_{\R^2} \frac{d^2k}{(2 \pi)^2}\: (k_j \gamma^j + m)\: \delta\big( k^2+m^2\big) \: \Theta(-k_0)\:
\hat{\psi}\big(k_1 \big) \: e^{\varepsilon k_0}\: e^{i k x}
\eeq
(for more details on this regularization method and the general context see~\cite[\S2.4.1]{cfs}
and~\cite[Section~1.2]{cfs}).
Next, for any~$x \in \scrM$ we define the {\em{local correlation operator}}~$F^\varepsilon$ by the
relations
\beq \label{Fepsdef}
\la u \,|\, F^\varepsilon(x)\, v \ra_\H 
= - \overline{({\mathfrak{R}}_\varepsilon \,u)(x)} ({\mathfrak{R}}_\varepsilon \,v)(x) \qquad \text{for all~$u,v \in \H$}\:.
\eeq
It is a bounded symmetric linear operator on~$\H$.
Taking into account that the inner product on the Dirac spinors at~$x$ has signature~$(2,2)$,
the local correlation operator~$F^\varepsilon(x)$ has rank at most four, and (counting multiplicities)
has at most two positive and at most two negative eigenvalues.

Varying the spacetime point~$x \in \scrM$, we obtain the so-called {\em{local correlation map}}
\beq \label{FeMink}
F^\varepsilon \::\: \scrM \rightarrow \F \:,
\eeq
where~$\F \subset \Lin(\H)$ denotes the set of all
symmetric operators on~$\H$ of finite rank which (counting multiplicities) have
at most one positive and at most one negative eigenvalue.
The last step is to drop all other structures (like the metric and causal structure
of Minkowski space, the spinorial structures, etc.).
Our concept is to work exclusively with the local correlation operators 
corresponding to the physical wave functions.
Thus the basic concept is that all spacetime structures (particles, fields, causal structure, geometry, \ldots) are encoded in the local correlation operators.
At this point it is obvious that this concept is sensible. But, as we shall see in the later sections in this book,
it is possible to reconstruct all spacetime structures from the local correlation operators.
In order to drop all the additional structures and to focus on the information contained in
the local correlation operators, we
introduce the measure~$\rho^\varepsilon$ on~$\F$ as the push-forward
of the volume measure on~$\scrM$
\beq \label{rhoMink}
\rho^\varepsilon := F^\varepsilon_* \mu
\eeq
(defined by~$\rho^\varepsilon(\Omega) := \mu ( (F^\varepsilon)^{-1}(\Omega))$,
where~$d\mu = dt \,dx$ is the two-dimensional volume measure on~$\scrM$).
We thus obtain a causal fermion system of spin dimension~$n=1$
(see Definition~\ref{defcfs}).

\subsection{The Fermionic Von Neumann Entropy} \label{secentropy2}
Having constructed the causal fer\-mion system, corresponding fermionic entropies
can be introduced as explained in Sections~\ref{secentropy}.
We now explain how and why these notions give us back the fermionic entanglement entropy
in Minkowski space as analyzed in~\cite{arealaw}.

The spacetime~$M$ of the causal fermion system is defined as the support of the measure~$\rho$
(see~\eqref{Mdef}). It turns out that this support coincides with the image of the
local correlation map~\eqref{FeMink} (the reason is that the mapping~$F^\varepsilon$ is
closed; for details see~\cite{oppio}). Moreover, using that the local correlation map is injective
(for details see again~\cite{oppio}), we  may
\beq \label{identify}
\text{identify~$x \in \scrM$ with~$F^\varepsilon(x) \in M$}\:,
\eeq
giving an identification of~$M$ with Minkowski space.
In this way, the spacetime pictures as shown in Figure~\ref{figfoliation} can be
translated to pictures in Minkowski space.
In particular, the set~$N_t \cap V$ can be associated to a spatial region in Minkowski space.

Next, we need to associate the commutator inner product~\eqref{OSIdyn} to the scalar
product in Minkowski space~\eqref{sprodMin}. To this end, we use the result from~\cite[Section~5.2]{noether}
that for Dirac wave functions which are {\em{macroscopic}} is much smaller than~$1/\varepsilon$,
these inner products coincide, i.e.,
\beq \label{spagree}
\la \psi | \phi \ra^t_\rho = c\: \la \psi | \phi \ra_m
\eeq
With this in mind, we choose the subspace~$\H^\fermi \subset \H$ 
(introduce before Definition~\ref{defSLrep}) as the span of all these macroscopic wave functions.
Then, since the scalar product on the right side of~\eqref{sigmadef} coincides by construction with
the restriction of~\eqref{sprodMin} to~$\H$, it follows that the operator~$\sigma$ is the identity,
\beq \label{sigmaone}
\sigma = \1 \:.
\eeq
As a consequence, the fermionic von Neumann entropy vanishes,
\[ S = 0 \:. \]
This is physically sensible, because the vacuum state is pure.

\subsection{The Fermionic Entanglement Entropy of a Causal Diamond} \label{secdiamone}
\begin{figure}
\psset{xunit=.5pt,yunit=.5pt,runit=.5pt}
\begin{pspicture}(251.00544607,225.44960322)
{
\newrgbcolor{curcolor}{0.80000001 0.80000001 0.80000001}
\pscustom[linestyle=none,fillstyle=solid,fillcolor=curcolor]
{
\newpath
\moveto(18.27661984,112.62883094)
\lineto(125.50273134,221.82610968)
\lineto(225.86058709,114.56064952)
\lineto(125.50273134,6.05324039)
\closepath
}
}
{
\newrgbcolor{curcolor}{0 0 0}
\pscustom[linewidth=0.99999871,linecolor=curcolor]
{
\newpath
\moveto(18.27661984,112.62883094)
\lineto(125.50273134,221.82610968)
\lineto(225.86058709,114.56064952)
\lineto(125.50273134,6.05324039)
\closepath
}
}
{
\newrgbcolor{curcolor}{0 0 0}
\pscustom[linewidth=2.4999988,linecolor=curcolor]
{
\newpath
\moveto(0.01077543,112.56283661)
\lineto(235.25604661,114.59081771)
}
}
{
\newrgbcolor{curcolor}{0 0 0}
\pscustom[linestyle=none,fillstyle=solid,fillcolor=curcolor]
{
\newpath
\moveto(231.69583565,121.56038291)
\lineto(251.00545383,114.72658877)
\lineto(231.81652104,107.56090982)
\curveto(235.28123969,111.66842771)(235.2310044,117.49570838)(231.69583565,121.56038291)
\closepath
}
}
{
\newrgbcolor{curcolor}{0 0 0}
\pscustom[linewidth=2.4999988,linecolor=curcolor]
{
\newpath
\moveto(19.27655055,0.01015)
\lineto(17.57166236,209.70013047)
}
}
{
\newrgbcolor{curcolor}{0 0 0}
\pscustom[linestyle=none,fillstyle=solid,fillcolor=curcolor]
{
\newpath
\moveto(10.60035294,206.14333611)
\lineto(17.443611,225.44960234)
\lineto(24.5998835,206.25715955)
\curveto(20.49406437,209.72389111)(14.66675977,209.6765121)(10.60035294,206.14333611)
\closepath
}
}
{
\newrgbcolor{curcolor}{0 0 0}
\pscustom[linewidth=1.75748033,linecolor=curcolor]
{
\newpath
\moveto(225.48306142,128.94234149)
\lineto(225.66474331,98.23634464)
}
\rput[bl](28,210){$t$}
\rput[bl](240,125){$x$}
\rput[bl](219,80){$\lambda$}
\rput[bl](118,91){$\Lambda$}
}
\end{pspicture}
\caption{A causal diamond.}
\label{figpos1}
\end{figure}
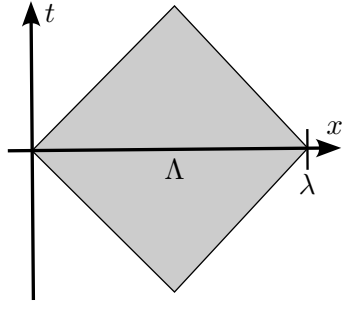

As an example for a causal fermion system with non-vanishing fermionic von Neumann entropy,
we now consider a flat {\em{causal diamond}} in two spacetime dimensions.
Given a closed interval~$\Lambda := (0, \lambda)$ with~$\lambda>0$, 
the corresponding spacetime~$(\scrD, g)$ is isometric to
the subset of two-dimensional Minkowski space
\beq \label{scrRdef}
\scrD = \big\{ (t,x) \in \scrM \;\;\;\text{with}\;\;\; x \in \Lambda \text{ and } |t| < \min(x, \lambda-x) \big\}\:;
\eeq
see Figure~\ref{figpos1}.
(see Figure~\ref{figpos1}).
Then the inclusions
\[ \scrD \subset \scrM \qquad \text{and} \qquad S\scrD = \scrD \times \C^2
\subset \scrM \times \C^2 = S\scrM \]
are clearly isometries.
The Dirac operator and the Dirac equation are given again by~\eqref{Dirop}
and~\eqref{Direq}.
On the level of the causal fermion system, the causal diamond is described in analogy to~\eqref{rhoMink}
by the measure
\beq \label{rhoepsDdef}
\rho^\varepsilon_\scrD := F^\varepsilon_* \mu_\scrD \qquad \text{with} \qquad \mu_\scrD = \chi_\scrD\, \mu \:.
\eeq
Using again the identification~\eqref{identify}, the diamond is described equivalently by the subset
\[ \scrD \subset \scrM \subset \F \qquad \text{and} \qquad \rho^\varepsilon_\scrD = \chi_\scrD\, \rho^\varepsilon \:. \]

The next step is to compute the localized
surface layer integral~\eqref{osiloc} and the localized one-particle density operator~$\sigma_V$ in~\eqref{sigmaVdef}. Here one must keep in mind that localizing the surface layer integral
is not the same as multiplying the wave functions by the characteristic function~$\chi_V$.
Indeed, the localization in~\eqref{osiloc} involves characteristic functions either for~$\psi$ or for~$\phi$,
but not for both wave functions. Due to the nonlocality of the surface layer integral, it makes a difference
where the characteristic function is inserted. For what follows, it suffices to note that the
nonlocality of the surface layer integral is on the Compton scale~$m^{-1}$. Therefore, this nonlocality
is negligible if the size of the region~$\lambda$ is much larger than~$m^{-1}$.
Under the assumption~$m \Lambda \ll 1$, instead of localizing the surface layer integral~\eqref{OSIdyn},
we may insert the localization directly into the spatial scalar product~\eqref{printM}.
This leads us to introducing the inner product
\beq \label{printD}
\la\psi | \phi\ra_{\scrD} := \int_{0}^\lambda \Sl \psi | \gamma^0 \phi \Sr|_{(0,x)}\: dx
\eeq
(note that, in contrast to~\eqref{printM}, we now integrate only over the interval~$(0, \lambda)$).
In this way, the localization to the spatial region~$V$ introduced in~\eqref{osiloc}
reduces to multiplication by a characteristic function, i.e.\ in analogy to~\eqref{sigmaVdef}
\[ 
\la\psi | \phi\ra_{\scrD} = \la u | \sigma_\scrD v \ra_\H \qquad \text{for all~$u,v \in \H^\fermi$} \]
with
\beq \label{sigmaVchi}
\sigma_\scrD = \chi_\Lambda \sigma \chi_\Lambda\:.
\eeq
Therefore, the fermionic entanglement entropy becomes
\beq \label{SVD}
S^\varepsilon_\scrD = \tr_{\H} \big(\eta(\sigma_\scrD) - \chi_\Lambda \,\eta(\sigma)\, \chi_\Lambda \big) \:.
\eeq

\subsection{An Area Law for the Entanglement Entropy of a Causal Diamond}
The entanglement entropy of~$\scrD \subset \scrM$ quantifies the entanglement between
the spacetime region~$\scrD$ with its causal complement. In our description of the quantum state
by the reduced one-particle operator, only those wave functions contribute which are non-zero
both in~$\scrD$ and in its complement. This is obvious in~\eqref{sigmaVchi} and~\eqref{SVD}
because the operator~$\eta(\sigma_\scrD) - \chi_\Lambda \,\eta(\sigma)\, \chi_\Lambda$ vanishes
on every vector which is a joint eigenstate of~$\sigma$ and~$\chi_V$, i.e.\ if
\[ \sigma \psi \sim \psi \qquad \text{and} \qquad \chi_V \psi = \psi \text{ or } 0\:. \]
In other words, the entanglement entropy is determined by those wave functions which are localized near the
boundary of the spatial region~$\Lambda$. This suggests that the entanglement entropy should scale
with the area of the boundary of~$\Lambda$. Such {\em{area laws}} have indeed
been proven in various situation with different techniques.
Clearly, for the two-dimensional diamond this boundary consists of the two points~$(0,0)$ and~$(0, \lambda)$
(see Figure~\ref{figpos1}). Therefore, in this case the area law simply states that the entanglement entropy
is non-zero and independent of the size~$\lambda$ of the diamond.

Another point of interest is the dependence of the entanglement entropy on the ultraviolet regularization.
For technical simplicity, we here consider the $i \varepsilon$-re\-gu\-la\-ri\-za\-tion~\eqref{Fourierreg}.
Then the dependence on the regularization is specified by the scaling behavior of the entanglement
entropy in~$\varepsilon$ for small~$\varepsilon>0$. For dimensional reasons, the
entanglement entropy should scale like~$\varepsilon^{-d+1}$, where~$d$ is the spatial dimension.
Moreover, an additional logarithmic singularity can arise, in which case we speak of an {\em{enhanced
area law}}. Typically, an enhanced area law is a result of long-range correlations, which are typical for massless
systems. In the present situation of one spatial dimension, we obtain an enhanced area law even in the
massive case. Here is our main result.

\begin{Thm} {\bf{(Area law for the entanglement entropy of a causal diamond)}} \label{thmdiamond}
The entanglement entropy of the causal diamond~\eqref{SVD} obeys the enhanced area law
\[ \lim_{\varepsilon \searrow 0} \:\frac{1}{ \log(1/\varepsilon)} S^\varepsilon_\scrD = \frac{1}{6}\:.\]
\end{Thm}
The proof of this theorem is given in~\cite{diamondentropy}.
We point out that this mathematical result holds both in the massive and massless case.
However, in the massless case the connection to causal fermion systems is debatable,
because it is no longer clear whether
the localized surface layer integral can again be approximated by~\eqref{printD}.

\section{Example: Four-Dimensional Minkowski Space} \label{secmink}
In this section we consider the example of a causal fermion system describing the Minkowski vacuum.
As we shall explain, this example gives the correspondence to the entanglement entropy
and the corresponding area law in~\cite{arealaw}. More precisely, we shall proceed in the
following two steps:
\bitem
\itemD Starting from the usual structures of relativistic quantum mechanics in Minkowski space,
we construct a causal fermion system describing the vacuum (Section~\ref{secmink4}).
\itemD We explain why and how specializing the notions of fermionic entropies introduced
in Sections~\ref{secentropy} and~\ref{secententropy} to this causal fermion system
gives back the usual notions of fermionic entropies as studied in~\cite{arealaw}
(Section~\ref{secmink2}).
\eitem
The purpose of these constructions is to give the fermionic entropies of a causal fermion
system a physical meaning by showing that they extend the well-established physical notions in
Minkowski space to a more general setting.

\subsection{Construction of the Causal Fermion System} \label{secmink4}
Proceeding similar as in Section~\ref{secmink1}, we now construct the causal fermion system
describing four-dimensional Minkowski space (more details can be found in~\cite[Section~1.2]{cfs} or~\cite[Chapter~5]{intro}). We let~$(\scrM, \la .,. \ra)$ be Minkowski space 
(with the signature convention~$(+ - - -)$) and~$d\mu$ the standard volume measure
(thus~$d\mu = d^4x$ in a reference frame~$x= (x^0, \ldots, x^3)$).
We consider the vacuum Dirac equation
\[ 
(i \gamma^j \partial_j - m) \psi = 0 \:, \]
where~$\gamma^j$ are the usual Dirac matrices in the Dirac representation, and~$m$ is the
rest mass (for simplicity of the presentation, we only consider one type of particles of mass~$m$;
the extension to several generations or systems involving leptons and quarks can be found in~\cite{cfs}).
Here the wave function~$\psi$ has four complex components, which describe the spinor components.
The spinors are endowed at each spacetime point with an inner product of signature~$(2,2)$,
which as in physics textbooks we denote by~$\overline{\psi} \phi$
(where~$\overline{\psi} := \psi^\dagger \gamma^0$ is the usual adjoint spinor).
For a solution~$\psi$ of the Dirac equation,
the function~$(\overline{\psi} \gamma^0 \psi)(t,\vec{x})$ has the interpretation as
the probability density of the Dirac particle at time~$t$ to be at the position~$\vec{x}$.
The spatial integral of this probability density is time independent as a consequence of the 
Dirac equation (conservation of the Dirac current).
Considering the bilinear form corresponding to this probability integral gives the
scalar product
\beq \label{sprodMin}
\la \psi | \phi \ra_m := \int_{\R^3} (\overline{\psi} \gamma^0 \phi)(t, \vec{x})\: d^3x \:.
\eeq
We denote the Hilbert space corresponding to this scalar product by~$\H_m = L^2(\R^3, \C^4)$;
it consists of all square-integrable wave function on~$\R^3$.
The Cauchy problem for the Dirac equation is well-posed, meaning that for
every square-integrable wave function at time~$t$ there is a corresponding global solution to the
Dirac equation. We usually identify the Cauchy data at time~$t$ with the corresponding solution.
In this way, the Hilbert space~$\H_m$ becomes the solution space of the Dirac equation.
On this solution space, the scalar product~\eqref{sprodMin} is
independent of time and also does not depend on the choice of the reference frame.

Similar as explained in~\eqref{Fourierneg} in two spacetime dimensions, we 
choose~$\H$ as the subspace of negative-frequency solutions. Moreover,
choosing a regularization operator~\eqref{Repsdef}, we form the local correlation operators~$F^\varepsilon(x)$
by~\eqref{Fepsdef}. In view of the signature of the spin inner product, these operators now have at most
two positive and at most two negative eigenvalues.
 Taking the push-forward of the resulting local correlation map (see~\eqref{FeMink} and~\eqref{rhoMink},
 we obtain a causal fermion system~$(\H, \F, \rho^\varepsilon)$ of spin dimension two.

\subsection{An Area Law for the Entanglement Entropy of a Spatial Subregion} \label{secmink2}
We now explain how the area law for the causal diamond in two spacetime dimensions
derived in Section~\ref{secdiamone} can be extended to four spacetime dimensions.
Exactly as explained in two spacetime dimensions in Section~\ref{secdiamone},
instead of the commutator inner product we consider the scalar product on a Cauchy surface~\eqref{sprodMin}
of Minkowski space. We let~$\Lambda \R^3$ be a bounded spatial subset, i.e.\ a subset of this Cauchy surface,
\[ \Lambda \subset \{t\} \times \R^3 \subset \scrM \:. \]
Moreover, for simplicity we choose the $i \varepsilon$-regularization, i.e.\ similar to~\eqref{Fourierreg}
we choose the regularization operator~\eqref{Repsdef} such that
\[ 
({\mathfrak{R}}_\varepsilon \psi)(t, \vec{x}) := \int_{\R^4} \frac{d^4k}{(2 \pi)^4}\: (k_j \gamma^j + m)\: \delta\big( k^2+m^2\big) \: \Theta(-k_0)\:
\hat{\psi}\big(k_1 \big) \: e^{\varepsilon k_0}\: e^{i k x} \]
(in~\cite{arealaw} more general regularizations with cutoff functions in momentum space are considered).
It is shown in~\cite[Theorem~1.1]{arealaw} that the entanglement entropy of the region~$\Lambda$
obeys the area law
\[ \lim_{\varepsilon \searrow 0} \varepsilon^2 \,S^\varepsilon_\Lambda = \mathfrak{M} \,  
\mathrm{vol}_2(\partial \Lambda) \:, \]
where~$\mathfrak{M}>0$ is a numerical constant.
We point out that this area law is {\em{not}} enhanced.
We remark that in~\cite{arealaw} this result was proven for more general regularizations
by cutoff functions. 
Moreover, the situation is considered that the region~$\Lambda$ is scaled by a parameter~$L$.
For simplicity, we here restrict attention to the special case of interest here.
In addition, corresponding area laws were proven for the R{\'e}nyi entropy.

We finally remark that this result for the
fermionic entanglement entropy can be used to {\em{define}} a notion of
two-dimensional area~$A$ for causal fermion systems, i.e.,
\[ A := c\, S_V \]
(with a constant~$c$ depending on the regularization length). An alternative, simpler method for defining
the two-dimensional area~$A$ is given by
\[ A := \int_{\partial \Omega_t \cap V} d\rho(x) \int_{M \setminus (\Omega_t \cup V)} d\rho(y)\: \L(x,y) \:. \]
(a similar definition was first given in~\cite{jacobson}).
Comparing these notions and deriving relations between them seems an interesting project for the future.

\section{Example: The Event Horizon of a Schwarzschild Black Hole} \label{secbh}
We now explain how to get a connection to the fermionic entanglement entropy of
a Schwarzschild black hole as studied in~\cite{bhentropy}.
The construction of the causal fermion system in Minkowski space
as outlined in Section~\ref{secmink} extends in a straightforward way to curved spacetime
(for details see~\cite[Section~1]{nrstg} or~\cite[Section~3.1]{grossmann}).
More precisely, in a globally hyperbolic spacetime, the solutions of the Dirac equation again
form a Hilbert space, where the scalar product is obtained similar to~\eqref{sprodMin}
by integrating the wave functions over a Cauchy surface. After introducing regularization operators~\eqref{Repsdef},
one can again introduce the local correlation operators by~\eqref{Fepsdef}
(where the product~$\overline{\psi} \phi$ is now the inner product of signature~$(2,2)$ on the
spinor space at~$x$). Taking the push-forward of the volume measure~$\mu = \sqrt{|\det g|}\, d^4x$
with respect to the local correlation map again~\eqref{rhoMink} gives the causal fermion system.

In the Schwarzschild black hole geometry, this construction can be carried out more explicitly
using the integral representation for the Dirac propagator as derived in~\cite{tkerr}, as we now outline.
In Schwarz\-schild coordinates, the line element of the Schwarzschild geometry takes the form
\[ 
ds^2 = g_{jk}\:dx^j \,dx^k = \Big( 1 - \frac{2M}{r} \Big) \: dt^2
- \Big( 1 - \frac{2M}{r} \Big)^{-1} \: dr^2 - r^2\: d \vartheta^2 - r^2\: \sin^2 \vartheta\: d\varphi^2\:, \]
where~$M>0$ is the mass of the black hole. 
We here restrict attention to the {\em{exterior region}}
outside the event horizon. Thus the coordinates~$(t,r, \vartheta, \varphi)$ are in the range
\[ -\infty<t<\infty,\;\;\; 2M <r<\infty,\;\;\; 0<\vartheta<\pi,\;\;\; 0<\varphi<2\pi \]
(here~$r=2M$ corresponds to the event horizon).
The exterior region is globally hyperbolic. The surfaces of constant coordinate time~$t$
form a foliation by Cauchy surfaces.
Similar to the Fourier representation in Minkowski space~\eqref{Fourierneg},
a general solution of the Dirac equation can be written as
\[ 
\psi(t,r,\vartheta, \varphi) = \sum_{k,n}
\int_{-\infty}^\infty d\omega \:e^{-i \omega t} \,\sum_{a=1}^2
\hat{\psi}_{a}^{kn}(\omega)\: \Psi^{k \omega n}_{m,a}(r,\vartheta,\varphi) \:, \]
where~$k \in \Z+1/2$ and~$n \in \N$ label the angular momentum modes, and~$\Psi^{k \omega n}_{m,a}(r,\vartheta,\varphi)$ are formed of fundamental solutions of the separated ODEs.
Noting that the integration variable~$\omega$ is the frequency of the solution,
we can choose the subspace~$\H$ of negative-frequency solutions simply as in~\eqref{Fourierneg}
by restricting the integration range to~$\omega \in (-\infty, 0]$. Moreover, the regularization
can be incorporated by inserting as in~\eqref{Fourierreg} an exponentially
decaying factor~$e^{\varepsilon \omega}$, i.e.,
\[ ({\mathfrak{R}}_\varepsilon \psi)(t,r,\vartheta, \varphi) := \sum_{k,n}
\int_{-\infty}^0 d\omega \:e^{\varepsilon \omega}\: e^{-i \omega t} \,\sum_{a=1}^2
\hat{\psi}_{a}^{kn}(\omega)\: \Psi^{k \omega n}_{m,a}(r,\vartheta,\varphi) \:. \]
Forming the local correlation operators and taking the push-forward of the volume measure
with respect to the local correlation map gives a causal fermion system describing the
exterior Schwarzschild geometry.

Working in Gaussian coordinates, one sees that the computations in Minkowski space in~\cite[Section~5]{noether}
also apply in curved spacetime. Therefore, similar to~\eqref{spagree}
the commutator inner product again coincides up to a prefactor with the scalar product.
Moreover, localizing the scalar product as in Figure~\ref{figfoliation} again reduces to multiplying
by a characteristic function~\eqref{sigmaVchi}. In this way, exactly as explained
for Minkowski space in Section~\ref{secmink}, the notions of fermionic entropies of the causal
fermion system reduce to the corresponding notions in quantum field theory.
In other words, the results in~\cite{bhentropy} extend to causal fermion systems, giving
a concise definition of the entanglement entropy of the event horizon. Moreover, this entanglement
entropy can be computed by counting the number of occupied angular momentum modes.
More precisely, the entanglement entropy can be written as the sum over the angular momentum modes
\[ S^\varepsilon = \sum_{k,n} S^\varepsilon_{k,n} \:, \]
and each mode satisfies an enhanced area law, i.e.\ similar to Theorem~\ref{thmdiamond},
\[ \lim_{\varepsilon \searrow 0} S^\varepsilon_{k,n} = \frac{1}{6}\:. \]
For the detailed definition of~$S^\varepsilon_{k,n}$ and the proof we refer to~\cite{bhentropy}.

\section{Example: Fermionic Lattices} \label{secspin}
In this section we illustrate how the  causal fermion system constructions above are related
to the usual treatment of entanglement and entropies in fermionic lattices in quantum many body physics and the K\"ahler structure formalism for quasi-free states (or synonymously Gaussian states).
In particular, we use this example to review the role of particle-number preserving states inside the larger class of quasi-free states.

It is natural to discuss this relation in the context of fermionic lattices, because it is typical for quantum many body models to have energy eigenstates which are not particle-number preserving, i.e., they are not of the type of quasi-free states we focus on here.
A prototypical example is the {\em{Kitaev chain}} model~\cite{alicea_new_2012, kitaev_unpaired_2001} 
\beq \label{Hkitaev}
H=-\mu \sum_{i=1}^N \Psi^\dagger_i\Psi_i - t\sum_{i=1}^N \left(\Psi_{i+1}^\dagger\Psi_i+ \Psi_i^\dagger \Psi_{i+1}\right)+\Delta \sum_{i=1}^N \left(\Psi_i^\dagger\Psi_{i+1}^\dagger+\Psi_{i+1}\Psi_i\right).
\eeq
If~$\Delta\neq0$ such that the so called \emph{pairing terms} or \emph{squeezing terms} do not vanish, the ground state and other eigenstates of~$H$, all of which are quasi-free states, have in general non-vanishing expectation values~$\left< \Psi_i\Psi_j\right>\neq0$.

The following subsection briefly reviews how to treat such general quasi-free states and how to calculate their entropies, also using the K\"ahler structure formalism (closely following~\cite{hackl-bianchi}).
We then discuss the particle-preserving quasi-free states with respect to a given total particle number operator, and finally give a causal fermion construction for  particle-preserving quasi-free lattice states.

\subsection{Review of Quasi-Free Fermionic States, Entropy and K\"ahler Structure}
Consider a lattice of~$N$ fermionic sites, such as the example of the one-dimensional Kitaev chain in~\eqref{Hkitaev}.
Every site~$i=1,\dots,N$ hosts one fermionic mode which has a creation operator~$\Psi_i^\dagger$ and an annihilation operator~$\Psi_i$  satisfying the canonical anti-commutation relations. We may organize the latter into one $2N\times 2N$-matrix,
\beq
G=\left(\begin{array}{c|c}\{ \Psi_i, \Psi_j \}  & \{ \Psi_i, \Psi_j^\dagger \} \\[0.3em] \hline \\[-1em] \{ \Psi_i^\dagger, \Psi_j \} & \{ \Psi_i^\dagger, \Psi_j^\dagger \}
\end{array}\right) = \left(\begin{array}{c|c} 0 & \1 \\ \hline \1 &0\end{array}\right).
\eeq
The creation and annihilation operators act on the $2^N$-dimensional quantum mechanical Hilbert space of fermionic
states denoted by~$\Fock$ which can be obtained by the usual Fock space construction: Starting from the joint vacuum state of all modes, characterized by~$\Psi_i \ket0=0$ for all~$i$, the Fock states are obtained by acting on the vacuum with all possible combinations of creation operators.

A {\em{thermal state}} of the Hamiltonian~$H$ has the statistical operator (or density matrix)
\[
W_\beta := \frac{1}{Z}\: \exp\big( -\beta H \big) \:, 
\]
where~$\beta = 1/(kT)$ is the inverse temperature and~$Z:=\tr_\Fock \exp\left({-\beta H}\right)$ is the partition function
(in the physics literature, this statistical operator is commonly denoted by~$\sigma_\beta$ or~$\rho_\beta$; we here use~$W_\beta$ in order to avoid confusion with the operators~$\sigma$ and~$\sigma_V$ in~\eqref{sigmadef} and~\eqref{sigmaVdef}, or the universal measure). 
If the Hamiltonian~$H$ is quadratic (as in~\eqref{Hkitaev}), then the thermal state is {\em{quasi-free}}.
In particular, in the limit of infinite inverse temperature~$\beta\to\infty$ we obtain the ground state of the Hamiltonian~$W_\beta\to W_\infty= \ket{g}\! \bra{g}$. 
Whereas the ground state is a pure state,  the thermal states at finite~$\beta$ are mixed, i.e.,  $W_\beta^2 \neq W_\beta$. 

In fact, all quasi-free  states can be obtained in this way: For every  mixed quasi-free state there exists a quadratic operator~$Q$ such that
\beq \label{Wbetagen}
W_\beta = \frac{1}{Z}\: \exp(-Q) \qquad \text{with} \qquad 
Z =\tr_\Fock \exp\left({-Q}\right)\:
\eeq
is its statistical operator. And if~$W=\ket{u}\!\bra{u}$ is a pure Gaussian state, then there exists a quadratic parent Hamiltonian whose ground state is~$\ket{u}$. (For more details, for example, see~\cite{hackl-bianchi}.)

Quasi-free states obey a Wick's theorem, meaning that higher order correlations can be calculated from the two-point correlation functions only. Hence, the density operator~$W_\beta$ acting on the fermionic Fock space~$\Fock$ (of~$2^N$ complex dimensions) is already fully characterized by the \emph{covariance matrix} 
\beq \label{Omegadef}
\Omega=(-\ii)\left(\begin{array}{c|c} \tr W_\beta (\Psi_i\Psi_j- \Psi_j\Psi_i) & \tr W_\beta (\Psi_i\Psi_j^\dagger- \Psi_j^\dagger\Psi_i)
\\[0.3em] \hline \\[-1em]
\tr W_\beta(\Psi^\dagger_i\Psi_j -\Psi_j\Psi^\dagger_i) & \tr W_\beta(\Psi_i^\dagger\Psi_j^\dagger-\Psi_j^\dagger\Psi_i^\dagger) \end{array}\right).
\eeq

The characterization of the state can be further compressed from this $2N\times2N$-matrix down to the~$N$ real numbers given by the occupation of the state's normal modes:
To this end, we diagonalize the generator~$Q$ of a given Gaussian state by a Bogoliubov transformation of the general form
\beq \label{psipdef}
\Psi'_i=\sum_{j=1}^n u_{ij}\Psi_j+v_{ij}\Psi_j^\dagger \:.
\eeq
In order to preserve the canonical anti-commutation relations, the coefficient matrices~$u_{ij}$
and~$v_{ij}$ must satisfy the relations
\beq \sum_{k=1}^N u_{ik} \,u^*_{jk} + v_{ik}\, v^*_{jk}
=\delta_{ij} \qquad \text{and} \qquad \sum_{k=1}^N u_{ik}v_{jk}+v_{ik}u_{jk}=0 \:. \label{eq:bogo_conditions}\eeq
By a suitable choice of these coefficient matrices, one can arrange that
the site operators~$\Psi_i$ are mapped the normal modes~$\Psi_i'$ of~$Q$, i.e.\
\beq \label{Qnormal}
Q= \sum_{i=1}^N  \beta_i\left(\Psi'^\dagger_i \Psi'_i-\Psi'_i\Psi'^\dagger_i\right)=\sum_{i=1}^N 2\beta_i  n_
i-\sum_{i=1}^N  \beta_i \:,
\eeq
where the~$n_i$ are the number operators~$n_i:=\Psi'^\dagger_i \Psi'_i$.
With respect to the normal modes, the state's density matrix is proportional to
$W_\beta \propto \exp(-\sum_i 2 \beta_i n_i)$, which means that it is given by a simple product state between the normal modes. More precisely, the partial state of each normal mode is given by
\[ W_i := \frac1{2\cosh \beta_i}\begin{pmatrix} \ee{- \beta_i}&0\\0& \ee{\beta_i}\end{pmatrix},
\]
and the normal modes are entirely uncorrelated from each other. This means that the total von Neumann entropy
is  the sum of the entropies of all normal modes, $S=\sum_{i=1}^N S_i$.
The only non-zero entries of the state's covariance matrix are
\[ \left< \Psi'_i{\Psi'}_i^\dagger-{\Psi'}_i^\dagger\Psi'_i\right> =\tanh \beta_i \:,\]
which is often written as~$\tanh \beta_i=\cos 2r_i$, with~$-\beta_i=\log\tan r_i$, to yield (see~\cite{hackl-bianchi})
\[ \left< {\Psi'}_i^\dagger\Psi'_i \right> =\frac12(1-\cos 2r_i)=\sin^2r_i \:. \]
Using this formula, one obtains the von Neumann entropy for each normal mode as
\[ S_i=-\tr W_i\log W_i= -\left(\left<{\Psi'}_i^\dagger\Psi'_i\right>\log \left<{\Psi'}_i^\dagger\Psi'_i\right>+ \left(1-\left<{\Psi'}_i^\dagger\Psi'_i\right>\right)\log\left(1-\left<{\Psi'}_i^\dagger\Psi'_i\right>\right)\right),
\]
in agreement with~\eqref{Sred} and~\eqref{etadef}.

Above we annotated the anti-commutation relations and the covariance matrix of the state as matrices in order to highlight the connection to the {\em{K\"ahler structure formalism}} for Gaussian states.
A strength of this formalism is that, in addition to illuminating the geometry of Gaussian states and phase space, it also allows us to treat fermionic and bosonic Gaussian states largely in parallel. 
Following the comprehensive review~\cite{hackl-bianchi}, we here briefly introduce the basic notions of the approach.

The matrices~$G$ and~$\Omega$ represent bilinear two-forms on the dual of the system's  phase space.
They can be contracted to yield a linear map on the system's phase space:
\beq \label{eq:Jdef}
J= \Omega G^{-1}=(-\ii)\left(\begin{array}{c|c}  \tr W_\beta (\Psi_i\Psi_j^\dagger- \Psi_j^\dagger\Psi_i) & \tr W_\beta (\Psi_i\Psi_j- \Psi_j\Psi_i)
\\[0.3em] \hline \\[-1em]
 \tr W_\beta(\Psi_i^\dagger\Psi_j^\dagger-\Psi_j^\dagger\Psi_i^\dagger) & \tr W_\beta(\Psi^\dagger_i\Psi_j -\Psi_j\Psi^\dagger_i) \end{array}\right).
\eeq
If the Gaussian state is pure, then~$J^2=-\id$ defines a linear complex structure on phase space, and~$(G,\Omega,J)$ define a K\"ahler structure on phase space.
If the state is not pure, then
\[ -1< J^2\leq 0, \]
and the eigenvalues of~$J$ are~$\pm\ii \cos 2r_i$ (in the notation above).
The von Neumann entropy of a  quasi-free state can then be expressed directly in terms of~$J$ as
\[ S=\left| \tr\left( \frac{\id+\ii J}2\log \left|\frac{\id+\ii J}2\right|\right)\right| . \]
(Note that this formula holds for both fermionic and bosonic Gaussian states~\cite{hackl-bianchi}.)

To close the subsection we discuss how to treat general subsystems.
In phase space, subsystems can be characterized by projection operators: If~$P$ is a projection operator acting on phase space, i.e.~$P=P^2$, which has even-dimensional rank and which is orthogonal with respect to~$G$, i.e.~$PG=GP^T$, then the image of~$P$, as a subspace of the total phase space, yields the phase space of a subsystem.
The state of the subsystem is then characterized by the restricted linear complex structure~$J_{|P}=PJP$.
A common and simple example of such a subsystem is, of course, a subset of~$v$ lattice sites. Here the restricted linear structure is the~$2v\times 2v$-matrix obtained by selecting the corresponding lines and columns of~$J$ in~\eqref{eq:Jdef}.

\subsection{Particle-Number Preserving States}
Whereas the previous subsection reviewed how to treat quasi-free states and their entropy in general, for the scope of this paper we restrict attention to particle-number preserving quasi-free states. 
As specified in Sec.~\ref{secentquasi}, in these states all squeezing terms~$\left<\Psi_i\Psi_j\right>=0$ vanish.
Hence the covariance matrix~$\Omega$ and the complex linear structure~$J$ both simplify. The latter, for example, is now given by
\beq
J=(-\ii) \left( \begin{array}{c|c} -2 D+\id _N &0\\ \hline 0 & 2D - \id_N
\end{array}\right),
\eeq
where~$(D)_{ij}=\left<\Psi_i^\dagger\Psi_j\right>$ is the state's reduced one-particle density operator.

Since~$D$ forms a Hermitian matrix, there exists a unitary Bogoliubov transformation from the lattice site modes~$\Psi_i$ to a set of normal modes~$\Psi'_i$  which diagonalizes~$D$. This Bogoliubov transformation has the form
\beq \label{eq:BogoZuPPNormalmodes}
\Psi_i'=\sum_j u_{ij}\Psi_j,
\eeq
which does not mix creation and annihilation operators. Since all~$v_{ij}=0$ vanish, it follows from~\eqref{eq:bogo_conditions} that the coefficients~$U=(u_{ij})$ form a unitary~$N\times N$-matrix.
For these modes we have 
\beq
\left<{\Psi'}^\dagger_i\Psi'_j\right>=\delta_{ij}\lambda_i
\eeq
and the reduced one-particle density operator reads
\beq \label{eq:D_as_Matrix}
D = U^T 
\mathrm{diag}(\lambda_1,\dots,\lambda_N)
 U^*,\text{ since }
\left<\Psi^\dagger_i\Psi_j\right>=\sum_{k,l} u_{ki} u^*_{lj}  \left<{\Psi'}^\dagger_k\Psi'_l\right>
= \sum_k u_{ki}u^*_{kj} \lambda_k
.
\eeq
Note that here~$0\leq\lambda_i\leq1$, whereas the normal modes used in the previous subsection have~$0\leq \lambda_i\leq\tfrac12$. (The latter condition could be achieved by following up the unitary Bogoliubov transformation with simple transformations of the type~$c_i\mapsto c_i^\dagger$.)

Since the Bogoliubov transformation between the lattice site modes~$\Psi_i$ and the normal modes~$\Psi_i'$ is unitary, the total number operator defined with respect to both sets of modes is identical,
\beq
\hat n= \sum_i n'_i = \sum_i {\Psi'}^\dagger \Psi'= \sum_{i,j,k} u_{ij}^* u_{ik} \Psi_j^\dagger\Psi_k = \sum_i \Psi_i^\dagger \Psi_i =\sum_in_i,
\eeq
which motivates the name \emph{particle-number preserving state}, together with the fact that the state's statistical density operator~$W$ commutes with~$\hat n$.

From this emerges a particularly clear interpretation of pure particle-number preserving quasi-free states:
Every such state is given by a simple product state of its normal modes in which a certain number of modes~$0\leq p\leq N$ is occupied.
For occupied modes we have~$\lambda_i=1$ whereas the not occupied modes have~$\lambda_i=0$.
In particular, every pure particle-number preserving quasi-free state is an eigenstate of the total number operator with~$\left<\hat n\right>=p$.

\subsection{Description in the Setting of Causal Fermion System}
In this subsection we construct a causal fermion system for a  particle-number preserving quasi-free state on a fermionic lattices and its subsystems, as discussed in the previous sections.

We begin with the case of a  pure state and use the notation from the previous subsection.
Assume that we are given a particle-number preserving quasi-free state of particle number~$p$.
Without loss of generality, we assume that the state is simply~$\psi'_1\wedge\dots\wedge\psi'_p$, i.e., the state where the first~$p$ normal modes are occupied.
Intuitively speaking, the Hilbert space~$\H$ of a causal fermion system can be thought of as the space spanned by all occupied fermionic states of the system.
Hence, we use~$\H=\C^p$ for the Hilbert space and denote its standard basis by~$\ket1,\dots,\ket p$.
On this Hilbert space each lattice site, labelled by~$k=1,\dots,N$, is represented by a so-called \emph{local correlation operator}~$F_k$  defined by
\beq
F_k= -\sum_{i,j=1}^p u^*_{ik} u_{jk} \ket i\bra j =\ket{x_k} \bra{x_k},\quad\text{with } \ket{x_k}=\sum_{i=1}^p u^*_{ik}\ket i.
\eeq
Clearly,~$F_k$ has rank (at most) one, and~$-\braket{x_k|x_k}$ is its only non-zero eigenvalue. 

The sum of all local correlation operators represents the reduced one-particle density operator. In fact, all operators have the property to sum up to the identity
\beq
\sum_{k=1}^N F_k=\sum_{i,j=1}^p\sum_{k=1}^N u^*_{ik}u_{jk}\ket i\bra j=\sum_{i,j=1}^p \delta_{ij}\ket i\bra j=\id_{\H}.
\eeq
Hence, this sum actually yields a representation of~$D$, if we view~$\H=\C^p\subset \C^N$ as a subset of the one-particle Hilbert space of the system, on which~$D$ acts as a projector onto the state's occupied modes.

More generally, in the case of a mixed particle-number preserving quasi-free state, $p$ is given by the rank of~$D$, i.e., the number of~$\lambda_i>0$ which are positive.
Then, with~$\H=\C^p$, we define the local correlation operators by
\beq
F_k=- \sum_{i,j=1}^p u^*_{ik} u_{jk}\sqrt{\lambda_i \lambda_j} \ket i\bra j =-\ket{x_k} \bra{x_k} \quad\text{with } \ket{x_k}=\sum_{i=1}^p u^*_{ik}\sqrt{ \lambda_i} \ket i.
\eeq
This operator is semi-negative definite and has rank at most one.
We denote the collection of these local correlation operators by
\[ M := \big\{ F_1, \ldots, F_N \} \subset \F \:, \]
where~$\F$ denotes the set of all negative semi-definite operators on~$\H$ of rank at most one.
Finally, on~$M$ we introduce the counting measure~$\rho$, i.e.,
\[ \rho(V) := \# (V \cap M) \:. \]
The resulting structure is very similar to that of a causal fermion system (see Definition~\ref{defcfs}).
The only difference is that~$\F$ is now formed of all symmetric operators which have zero positive
and at most one negative eigenvalues. Using a notion first introduced in~\cite[Definition~2.5]{topology},
the triple~$(\H, \F, \rho)$ is a {\em{Riemannian fermion system}} of spin dimension one.
Riemannian fermion systems can be understood as a Euclidean variant of a causal fermion system.

Also in the mixed state construction, the sum of all local correlation operators  yields a representation of the reduced one-particle density operator. In terms of the measure~$\rho$, we  can define the operator
\beq\label{eq:sigmalattice}
\sigma=\int_M d \rho =\sum_{k=1}^N F_k = \sum_{i=1}^p \lambda_i \ket i \bra i.
\eeq
This operator is isospectral to~$D$, apart from zero eigenvalues which do not contribute to entropies. (Again as above, we may also view~$\H\subset\C^N$ as a subspace of the one-particle space such that~$\sigma$ is basically equal to~$D$.)

In order to  capture a subsystem given by a subset of lattice sites~$V\subset M$ we simply restrict the integration in~\eqref{eq:sigmalattice} to~$V$ and define the operator
\beq
\sigma_V=\int_V d \rho =\sum_{k\in V} F_k.
\eeq
If for  simplicity  and without loss of generality, we assume that~$V=\{1,\dots, v\}$, then
\beq\label{eq:sum_for_sigmaV_lattice}
\sigma_V=\sum_{k=1}^v \sum_{i,j=1}^p u^*_{ik}u_{jk}\sqrt{ \lambda_i \lambda_j}\ket i \bra j=\sum_{k\in V}\ket{x_k}\bra{x_k}.
\eeq
This operator is, apart from zero eigenvalues,  isospectral  to the reduced one-particle density operator~$D_V$ of the subsystem~$V$.
To see this, first note that~$D_V$ corresponds to first $v\times v$-diagonal block of the~$N\times N$-matrix~$D$ in~\eqref{eq:D_as_Matrix}. Hence, if we take~$U_V$ to be the~$N\times v$ matrix consisting of the first~$n$ columns of~$U$, then
\beq
D_V= (U_V)^T \mathrm{diag}(\lambda_1,\dots,\lambda_N)U_V^*.
\eeq
This matrix is isospectral, apart from zero eigenvalues, to the $N\times N$-matrix 
\begin{equation}
 \mathrm{diag}(\sqrt{\lambda_1},\dots,\sqrt{\lambda_N})U_V^* (U_V)^T \mathrm{diag}(\sqrt{\lambda_1},\dots,\sqrt{\lambda_N}).
\end{equation}
However, all matrix entries of this matrix outside of its first $p\times p$-diagonal block vanish, since~$\lambda_{p+1}=\dots=\lambda_N=0$.
This block in turn is exactly the matrix representation of~$\sigma_V$, as seen from~\eqref{eq:sum_for_sigmaV_lattice}.

The above shows that the system's state, and the marginal states of subsystems are captured respectively captured by~$\sigma$ and~$\sigma_V$ equivalently to the reduced one-particle density operators.
In particular, we  obtain the von Neumann entropy of the total system as~$S=\tr_{\H}(\eta(\sigma))$ and of a subsystem as 
\beq S_V=\tr_{\H}(\eta(\sigma_V)).
\eeq

We highlight that the subsystems obtained by restricting the measure~$\rho$ to a subset of~$M$ do not exhaust all possible subsystems which can be defined in terms of projection operators on phase space as discussed above. The presented construction only captures subsystems whose corresponding projection operator is diagonal with respect to lattice site modes.
In particular, it does not apply to potentially more abstract subsystems where the diagonalization of the projection operator requires a general Bogoliubov transformation that mixes annihilation and creation operators.

Instead the construction is apt for scenarios in which the reference basis modes, here the lattice site modes, and the state's eigenmodes agree on the notion of particle numbers. As a consequence for pure particle-number preserving quasi-free states non-trivial entanglement entropy of subsystems only arise if the state is neither the vacuum state of no particles or the fully occupied state.
In these two cases, because all~$\ket{x_k}$ have either unit norm or vanish, for all subsets~$V\subset M$ the only eigenvalues of~$\sigma_V$ are either zero or one, such that~$S(\sigma_V)=0$.
(In the standard formalism this corresponds to~$D=0$ or~$D=\id$ such that also the restrictions to orthogonal subspaces either vanish or are equal to the identity.)

\section{The Fermionic Relative Entropy} \label{secrelative}
We now consider two causal fermion systems~$(\H, \F, \rho)$ (describing the vacuum)
and~$(\tilde{\H}, \tilde{\F}, \tilde{\rho})$ (describing the interacting system).
In order to relate the two systems to each other, we assume that both Hilbert spaces
come with isometric embeddings into a ``larger'' Hilbert space~$\H^\text{tot}$, i.e.,
\[ \iota_\H : \H \hookrightarrow \H^\text{tot} \:, \qquad \iota_{\tilde{\H}} : \tilde{\H} \hookrightarrow \H^\text{tot} \:. \]
This also gives a corresponding embedding of~$\F$ into~$\F^\text{tot}$,
\[ \iota_\F : \F \hookrightarrow \F^\text{tot} \:, \qquad x \mapsto \iota_\H \circ x \circ \pi_{\iota(\H)} \]
(where~$\pi_{\iota(\H)} : \H^\text{tot} \rightarrow \iota(\H) \subset \H^\text{tot}$ is the orthogonal projection).
Indeed, given an operator~$x \in \F$, the operator~$\iota_\F x \in \Lin(\H^\text{tot})$
has again at most~$n$ positive and at most~$n$ negative eigenvalues and is therefore in~$\F^\text{tot}$.
Similarly, we introduce the embedding~$\iota_{\tilde{\F}} : \tilde{\F} \hookrightarrow \F^\text{tot}$.
Finally, we introduce the measures~$\iota_* \rho$ and~$\tilde{\iota}_* \tilde{\rho}$ on~$\F^\text{tot}$ by
taking the push-forward. In this way, both spacetimes are described on the same Hilbert space~$\H^\text{tot}$.
For notational convenience, we usually omit the embeddings.

Choosing past sets~$\Omega^t \subset M$ and~$\tilde{\Omega}^t \subset \tilde{M}$,
the relation~\eqref{sigmadef} defines two operators~$\sigma$ and~$\tilde{\sigma}$ on~$\H^\text{tot}$.
\begin{Def} \label{defrelative}
The {\bf{relative entropy}} of~$\tilde{\sigma}$ with respect to~$\sigma$ is defined by
\beq \label{Sreldef}
S^\rel(\tilde{\sigma}, \sigma)
= -\tr_{\H^\text{\rm{tot}}} \Big(  \tilde{\sigma} \big( \log \tilde{\sigma} - \log \sigma \big)
+ (\1-\tilde{\sigma})\, \big( \log( \1 - \tilde{\sigma}) - \log( \1 - \sigma) \big) \Big) \:,
\eeq
where~$\eta$ is again the von Neumann entropy function~\eqref{etadef}.
\end{Def} \noindent
This formula is derived in Theorem~\ref{thmrelreduced} in Appendix~\ref{appA}.

The relative entropy can be applied in various ways. One situation is that one causal fermion
system is a subsystem of the other. To this end, given a causal fermion system~$(\H, \F, \rho)$, we choose a closed
subset of subset~$\tilde{M} \subset M$ and introduce the measure
\[ \tilde{\rho} = \chi_{\tilde{M}} \rho \:. \]
Then~$(\H, \F, \tilde{\rho})$ defines a causal fermion system describing the subregion~$\tilde{M}$.

As another application, one can consider two causal fermion systems
in the same classical spacetime. To this end, given a globally hyperbolic spacetime~$(\scrM, g)$, we
choose a Cauchy surface~$\scrN$ and the Hilbert space~$\H^\text{tot} = L^2(\scrN, S\scrM)$ with scalar product
\[ \la \psi | \phi \ra_{\H^\text{tot}} := \int_\scrN \Sl \psi | \slashed{\nu} \, \phi \Sr(x)\: d\mu_{\scrN}(x) \]
(where~$\slashed{\nu}$ denotes Clifford multiplication by the future-directed normal~$\nu$,
and~$\mu_\scrN$ is the volume measure on~$\scrN$ of the induced Riemannian metric).
Solving the Cauchy problem, this Hilbert space can be identified with a space of
weak solutions of the Dirac equation. 
We choose~$\H, \tilde{\H} \subset \H^\text{tot}$ as subspaces of this solution space.
Moreover, we choose~$\,\,\tilde{\!\!\scrM}$ as~$\scrM$ or a globally hyperbolic
subset of~$\scrM$. Finally, we construct~$\rho$ and~$\tilde{\rho}$ as the push-forward of the local
correlation map (for details see~\cite{nrstg}).

We now illustrate the last construction in two simple examples which tie in to the examples
in two-dimensional spacetimes considered in Section~\ref{secminktwo}.

\subsection{Example: Finite Particle Systems in Two-Dimensional Minkowski Space} \label{secrelmink}
We again consider the Dirac equation in two-dimensional Minkowski space.
We denote the Hilbert space of all Dirac solutions with the scalar product~\eqref{printM}
again by~$(\H_m, (.|.)_m)$. Slightly generalizing the construction in Section~\ref{secmink1},
we now introduce the regularization operators~$({\mathfrak{R}}_\varepsilon)_\varepsilon$
on all of~$\H_m$, i.e.\ in modification of~\eqref{Repsdef}
\[ 
{\mathfrak{R}}_\varepsilon \::\: \H_m \rightarrow C^0(\scrM, S\scrM) \:. \]
As a concrete example one can consider the mollification by convolution with with a test function, i.e.\
\[ ({\mathfrak{R}_\varepsilon\, \psi}_\varepsilon)(t,x) := \frac{1}{\varepsilon} \int_{\infty}^\infty
\eta\Big(\frac{x-y}{\varepsilon} \Big)\: \psi(t, y)\: dy \]
with~$\eta \in C^\infty_0(\R, \R)$. The precise choice of the regularization operators is irrelevant
because, as we shall see below, the relative entropy will be well-defined even in the limit~$\varepsilon \searrow 0$
when the regularization is removed.

We choose~$\H \subset \H_m$ exactly as in Section~\ref{secmink1} as the subspace of all
negative-frequency solutions of the Dirac equation. Forming the local correlation map~\eqref{FeMink}
and taking the push-forward of the volume measure~\eqref{rhoMink} gives the causal fermion
system~$(\H, \F, \rho^\varepsilon)$
describing the vacuum. Next we let~$\tilde{\H} \subset \H_m$ be another subspace with the property
that it differs from~$\H$ only on a finite-dimensional subspace, i.e.\
\beq \label{finitedim}
\H^\perp \cap \tilde{\H} \quad\text{and} \quad \tilde{\H}^\perp \cap \H \qquad \text{are finite-dimensional subspaces of~$\H_m$}\:.
\eeq
Forming the corresponding local correlation map~$\tilde{F}^\varepsilon$
and taking its push-forward of the volume measure gives the causal fermion
system~$(\tilde{\H}, \tilde{\F}, \tilde{\rho}^\varepsilon)$.

We now explain how to compute the relative entropy~\eqref{Sreldef}.
As explained in Section~\ref{secentropy2}, we choose the subspaces~$\H^\fermi \subset \H$
and~$\tilde{\H}^\fermi \subset \tilde{\H}$ as all the negative energy solutions
which are macroscopic in the sense that their energy is much smaller than~$1/\varepsilon$.
Then, similar to~\eqref{sigmaone}, the operators~$\sigma$ and~$\tilde{\sigma}$ are the identity
on their respective Hilbert spaces. Considering them as operators on the whole solution space~$\H^\text{tot}:=\H_m$, they are projection operators to the subspaces~$\H^\fermi$ and~$\tilde{\H}^\fermi$,
respectively, i.e.\
\[ \sigma = \pi_{\H^\fermi}\:,\tilde{\sigma} = \pi_{\tilde{\H}^\fermi} \::\: \H^\text{tot} \rightarrow \H^\text{tot} \:. \]
Using these formulas in~\eqref{Sreldef}, the operators~$\log \tilde{\sigma} - \log \sigma$
and~$\log( \1 - \tilde{\sigma}) - \log( \1 - \sigma)$ have rank bounded by the dimensions
of the finite-dimensional subspaces~\eqref{finitedim}. Therefore, the trace is well-defined
(it may be infinite in case that the operators~$\sigma$ or~$\1-\sigma$ have a non-trivial kernel
on the subspaces~$\tilde{\H}^\fermi$ respectively~$(\tilde{\H}^\fermi)^\perp$).
Moreover, one sees that the relative entropy remains well-defined in the limit when~$\H^\fermi$
exhausts~$\H$ and~$\tilde{H}^\fermi$ exhausts~$\tilde{H}$. Moreover, taking the
limit~$\varepsilon \searrow 0$ does not give rise to divergences.
In this way, the relative entropy is well-defined even without ultraviolet cutoff.

\subsection{Example: Finite Particle Systems in a Causal Diamond} \label{secreldiamond}
As a modification of the previous example, we now restrict the systems in Minkowski space
to the causal diamond. Thus, following the procedure in Section~\ref{secdiamone}, we
denote the volume measure in the diamond~$\scrD$~\eqref{scrRdef} by~$\mu_\scrD$
and similar to~\eqref{rhoepsDdef} and introduce the measures
\[ \rho^\varepsilon_\scrD := F^\varepsilon_* \mu_\scrD \qquad \text{and} \qquad
\tilde{\rho}^\varepsilon_\scrD := \tilde{F}^\varepsilon_* \mu_\scrD \:. \]
In this way, we obtain two causal fermion systems~$(\H, \F, \rho^\varepsilon_\scrD)$
and~$(\tilde{\H}, \tilde{\F}, \tilde{\rho}^\varepsilon_\scrD)$ describe the causal diamond
in the vacuum and containing a finite number of particles and anti-particles.
Similar to~\eqref{sigmaVchi}, the corresponding one-particle density operators are obtained
from those in Minkowski space by multiplying with characteristic functions, i.e.\
\[ \sigma_\scrD = \chi_\Lambda \sigma \chi_\Lambda \qquad \text{and} \qquad
\tilde{\sigma}_\scrD = \chi_\Lambda \tilde{\sigma} \chi_\Lambda \:. \]
Now we can compute the relative entropy by~\eqref{Sreldef} (adding a subscript~$\scrD$
to all operators~$\sigma$ and~$\tilde{\sigma}$). This entropy can be understood as the
{\em{relative fermionic entanglement entropy}} of the diamond.

\appendix
\section{Expressing Fermionic Entropies in Terms of the Reduced One-Particle Density Operator} \label{appA}
In this appendix, we explain how the von Neumann entropy of a quasi-free Fermi gas
can be expressed in terms of the reduced one-particle density operator.
We assume throughout that the state preserves the particle number.
The formula derived in Theorem~\ref{thmreduced} appears commonly in the literature
(see for example~\cite[Equation 6.3]{ohya-petz}, \cite{klich, casini-huerta, longo-xu}
and~\cite[eq.~(34)]{helling-leschke-spitzer}). For completeness, in this appendix
we give detailed proofs, which might be of independent interest.
The method of proof is by direct computation. However, the computation is not quite straightforward.
We are grateful to Wolfgang Spitzer for explaining us the basic steps of the computation.

For technical simplicity, we restrict attention to the finite-dimensional setting.
The resulting formula can be extended in a straightforward way to infinite dimensions
by choosing an exhaustion and taking the limit, provided that all appearing operators are trace-class.
Thus we let~$(\H, \la .|. \ra_\H$) be the one-particle Hilbert space, which we assume to be finite-dimensional.

\begin{Thm} \label{thmreduced} The von Neumann entropy~$S$ of the quasi-free state~$\omega$ 
(as defined by~\eqref{S1}) can be expressed in terms of the reduced one-particle density operator~$D$
(as defined in~\eqref{Ddef} and~\eqref{omega2def}) by~\eqref{S2}.

Moreover, given a parameter~$\varkappa \in \R^+ \setminus \{1\}$, the R{\'e}nyi entropy defined by
\beq \label{Skappafock}
S_\varkappa(W) := \frac{1}{1-\varkappa} \log \tr_\Fock \big( W^\varkappa \big)
\eeq
can be written as
\[ S_\varkappa = -\tr_\H \eta_\varkappa(D) \:, \]
with
\[ \eta_\varkappa(t) := \frac{1}{1-\varkappa}\,\log \big( t^\varkappa + (1-t)^\varkappa \big) \:. \]
\end{Thm}

We now enter the proof of this theorem, which will be completed after the proof of Lemma~\ref{lemmatrace}.
We first explain how the statistical operator~$W$ can be constructed from a given
reduced one-particle density operator. Thus let~$D$ be a symmetric operator with~$0 \leq D \leq \1$.
We choose an orthonormal eigenvector basis~$e_1, \ldots, e_N$ of~$D$, i.e.\
\[ D e_n = d_n \, e_n \qquad \text{with} \qquad 0 \leq d_n \leq 1 \:. \]
In preparation of the general construction of~$W$,
we begin with the case~$N=1$ of a one-dimensional Hilbert space.
In this case, the Fock space~$\Fock$ is two-dimensional, spanned by the vacuum~$\ket0$ and the
Fock vector~$\Psi^\dagger \ket0$ where the one-particle state~$e_1$ is occupied, i.e.\
\[ \Fock = \text{span} \big( \ket0, \Psi^\dagger \ket0 \big) \:. \]
Using a matrix notation, the state~$\omega$ is represented by an operator~$W$ on~$\F$ having the
matrix representation
\beq \label{Wmatrix}
W = \begin{pmatrix} 1-d & 0 \\ 0 & d \end{pmatrix}\:,
\eeq
where we set~$d=d_1$.
This is a statistical operator, where the one-particle state~$e_1$ is occupied with probability~$d$.
Decomposing it into a convex combination of pure states,
\[ W = \big(1-d) \: | 0 \ra \la 0 | + d\, | \Psi^\dagger | 0 \ra \la 0 | \Psi | \:, \]
each summand is obviously a quasi-free, and so is~$W$.
For what follows, it is preferable to rewrite~$W$ as
\begin{align}
W &= \big(1-d \big) \: | 0 \ra \la 0 | + d\, \Psi^\dagger\Psi = \big(1-d \big) \:\Big(  | 0 \ra \la 0 | + \frac{d}{1-d}\, \Psi^\dagger\Psi \Big) \notag \\
&= \big(1-d \big) \:\Big(  | 0 \ra \la 0 | + e^{-s}\, \Psi^\dagger\Psi \Big) \quad \text{with} \quad
s := \log \Big( \frac{1-d}{d} \Big) \:. \label{Wprod}
\end{align}
This operator can be rewritten as
\[ W = \big(1-d \big) \:e^{-s \Psi^\dagger \Psi} \:, \]
as is immediately verified by writing out the matrix representation~\eqref{Wmatrix}.

In the case~$N>1$ of general dimension, the Fock state has a similar structure, as is made precise in the next lemma.
We denote the creation and annihilation operator corresponding to the basis~$e_n$ by~$\Psi_n^\dagger$
and~$\Psi_n$, respectively.
\begin{Lemma} \label{lemmaWdef}
The quasi-free state~$\omega$ having the one-particle density~$D$ can be
represented on the Fock space~$\Fock$ by
\beq \label{statedef}
\omega(A) = \tr_{\Fock} \big( W A \big) \:,
\eeq
where
\beq \label{Wdef}
W := \det(\1-D) \:\exp \bigg( -\sum_{n=1}^N s_n\, \Psi_n^\dagger \Psi_n \bigg)
\eeq
and
\beq \label{sndef}
s_n := \log \Big( \frac{1-d_n}{d_n} \Big)\:.
\eeq
\end{Lemma}
\Proof We first verify by direct computation that~$W$ is a density operator.
It is obviously positive. Therefore, it remains to show that it has trace one.
The operator~$\Psi_n^\dagger \Psi_n$ has the eigenvalues zero and one,
depending on whether the state~$e_n$ is occupied or not. Therefore, the operator
\[ e^{-s_n \Psi_n^\dagger \Psi_n} \qquad \text{has the eigenvalues~$1$ and~$e^{-s_n}$} \:. \]
Similarly, the operator
\beq \label{evalex}
\exp \bigg( -\sum_{n=1}^N s_n\, \Psi_n^\dagger \Psi_n \bigg)
\qquad \text{has the eigenvalues} \qquad \prod_{n=1}^N e^{-p_n s_n} \;\; \text{with} \;\;
p_n \in \{0,1\} \:.
\eeq
Taking the trace gives
\begin{align*}
&\tr_\Fock \exp \bigg( -\sum_{n=1}^N s_n\, \Psi_n^\dagger \Psi_n \bigg)
= \sum_{p_1, \ldots, p_N \in \{0,1\}} \;\;\prod_{n=1}^N e^{-p_n s_n} \\
&= \prod_{n=1}^N \sum_{p_n \in \{0,1\}} e^{-p_n s_n}
= \prod_{n=1}^N \Big(1 + e^{-s_n} \Big) 
\overset{\eqref{sndef}}{=} \prod_{n=1}^N \Big( 1 + \frac{d_n}{1-d_n} \Big)
= \prod_{n=1}^N \frac{1}{1-d_n} \:.
\end{align*}
Therefore,
\begin{align*}
\tr_\Fock W = \det(\1-D) \;\tr_\Fock \exp \bigg( -\sum_{n=1}^N s_n\, \Psi_n^\dagger \Psi_n \bigg)
= \prod_{m=1}^N (1-d_m)\: \prod_{n=1}^N \frac{1}{1-d_n} = 1\:,
\end{align*}
concluding the proof that~$W$ is a density operator.

Let us verify that the state defined by~\eqref{statedef} is quasi-free.
This can be seen in various ways. One method is to write~\eqref{sndef} similar to~\eqref{Wprod} as
\[ W = \det(\1-D) \prod_{n=1}^N \Big( \1 + e^{-s_n}\, \Psi^\dagger_n \Psi_n \Big) \:. \]
Multiplying out, we get a sum of terms. Each summand describes a pure product state,
which clearly is quasi-free. Taking a convex combination, it follows that also~$W$ is quasi-free.

We finally verify that reduced one-particle density of~$W$ coincides with~$D$.
The two-point function is computed as follows.
\begin{align*}
\tr_\Fock \big( \Psi_k^\dagger \Psi_k\: W \big) &=
\det(\1-D) \:\tr_\Fock \Big\{ \Psi_k^\dagger \Psi_k\: e^{-\sum_{n=1}^N s_n\, \Psi_n^\dagger \Psi_n} \Big\} \:.
\end{align*}
Similar to~\eqref{evalex}, the operator inside the curly brackets has the eigenvalues
\[ 
p_k \,e^{-s_k} \prod_{n=1}^N e^{-p_n s_n} \:. \]
Hence
\beq \label{focktrace}
\tr_\Fock \bigg\{ \Psi_k^\dagger \Psi_k\: e^{-\sum_{n=1}^N s_n \Psi_n^\dagger \Psi_n} \bigg\}
= e^{-s_k} \prod_{n \neq k} \Big(1 + e^{-s_n} \Big) = e^{-s_k} \prod_{n \neq k} \frac{1}{1-d_n} \:,
\eeq
and thus
\[ \tr_\Fock \big( \Psi_k^\dagger \Psi_k\: W \big) = (1-d_k)\: e^{-s_k}
\overset{\eqref{sndef}}{=} (1-d_k)\: \frac{d_k}{1-d_k} = d_k \:. \]
This concludes the proof.
\QED

We next compute the von Neumann entropy of~$W$.
\begin{Lemma} \label{lemmatrace}
\[ S(W) = -\tr_{\H} \big( D \log D + (\1-D)\, \log(\1-D) \big) \:. \]
\end{Lemma}
\Proof From~\eqref{Wdef} it follows that
\[ W \log W = W \log \det(\1-D) - W \sum_{n=1}^N s_n\, \Psi_n^\dagger \Psi_n \:. \]
Taking the trace and using that~$W$ has trace one, we obtain
\begin{align*}
S &= -\tr_\Fock \big(W \log W) \\
&=-\log \det(\1-D) 
+\det(\1-D) \sum_{k=1}^N s_k\, \tr_\Fock \bigg\{ \Psi_k^\dagger \Psi_k\: e^{-\sum_{n=1}^N s_n \Psi_n^\dagger \Psi_n} \bigg\}\:.
\end{align*}
Using~\eqref{focktrace}, we get
\begin{align*}
&\tr_\Fock \big(W \log W) = \log \det(\1-D) 
- \det(\1-D) \sum_{k=1}^N s_k\, e^{-s_k} \prod_{n \neq k} \frac{1}{1-d_n} \\
&= \log \det(\1-D) -\sum_{k=1}^N (1-d_k)\: s_k\, e^{-s_k} \\
&= \log \det(\1-D) -\sum_{k=1}^N (1-d_k) \log \Big( \frac{1-d_k}{d_k} \Big)\: \frac{d_k}{1-d_k} \\
&= \sum_{k=1}^N \log \big( 1-d_k \big) -\sum_{k=1}^N d_k \log \Big( \frac{1-d_k}{d_k} \Big) \\
&= \sum_{k=1}^N \log \big( 1-d_k \big) -\sum_{k=1}^N d_k \log \big( 1-d_k \big) 
+ \sum_{k=1}^N d_k \log \big( d_k \big) \\
&= \sum_{k=1}^N d_k \log d_k + \sum_{k=1}^N (1-d_k) \log \big( 1-d_k \big) \\
&= \tr_{\H} \big( D \log D + (\1-D)\, \log(\1-D) \big) \:.
\end{align*}
This gives the result.
\QED
Combining Lemmas~\ref{lemmaWdef} and~\ref{lemmatrace}, 
one obtains the statement of Theorem~\ref{thmreduced} for the von Neumann entropy.
It remains to generalize this result to the R{\'e}nyi entropy:
\begin{Lemma} The R{\'e}nyi entropy of~$W$ can be expressed by
\[ S_\varkappa(W) = \tr_{\H} \eta_\varkappa(D) \:. \]
\end{Lemma}
\Proof From~\eqref{Wdef} it follows that
\begin{align*}
W^\varkappa &= \det\nolimits^\varkappa(\1-D) \:\exp \bigg( - \varkappa \sum_{n=1}^N s_n\, \Psi_n^\dagger \Psi_n \bigg) \\
\tr_\Fock\big( W^\varkappa \big) &= \det\nolimits^\varkappa(\1-D) \tr_\Fock \bigg\{
\exp \bigg( - \varkappa \sum_{n=1}^N s_n\, \Psi_n^\dagger \Psi_n \bigg) \bigg\} \\
&= \det\nolimits^\varkappa(\1-D) \tr_\Fock \bigg\{ \prod_{n=1}^N
\exp \bigg( - \varkappa s_n \Psi_n^\dagger \Psi_n \bigg) \bigg\} \\
&= \prod_{n=1}^N (1-d_n)^\varkappa\: 
\tr_{\Fock_n} \Big\{  e^{-\varkappa s_n \Psi_n^\dagger \Psi_n} \Big\} \:,
\end{align*}
where~$\Fock_n$ is the two-dimensional Fock space generated from the vacuum by acting with the
operator~$\Psi_n^\dagger$. Choosing a matrix representation similar to~\eqref{Wmatrix}, we obtain
\[ e^{- \varkappa \Psi_n^\dagger \Psi_n} = \begin{pmatrix} 1 & 0 \\ 0 & e^{-\varkappa s_n} \end{pmatrix}\:. \]
Hence
\[ \tr_{\Fock_n} \Big\{  e^{-\varkappa s_n \Psi_n^\dagger \Psi_n} \Big\} =
1 + e^{-\varkappa s_n} \overset{\eqref{sndef}}{=} 1 + \Big( \frac{d_n}{1-d_n} \Big)^\varkappa \:. \]
We conclude that
\[ W^\varkappa = \prod_{n=1}^N \Big( \big(1-d_n \big)^\varkappa + d_n^\varkappa \Big)  \:. \]
Using this formula in~\eqref{Skappafock} yields
\[ S_\varkappa(W) = \frac{1}{1-\varkappa} \sum_{n=1}^N \log \Big( \big(1-d_n \big)^\varkappa + d_n^\varkappa \Big)
= \tr_{\H} \eta_\varkappa(D) \:, \]
giving the result.
\QED
This completes the proof of Theorem~\ref{thmreduced}

We next prove a corresponding result for the {\em{entanglement entropy}}.
To this end, we begin with a subspace~$\H_A \subset \H$ of the one-particle Hilbert space.
The fermionic Fock space generated by~$\H_A$ is denoted by~$\Fock_A$. It is clearly a subspace
of~$\Fock$. Moreover, choosing an orthogonal decomposition
\[ \H = \H_A \oplus \H_B \:, \]
we get a corresponding tensor product structure for the Fock spaces,
\[ \Fock = \Fock_A \otimes \Fock_B \:. \]
Now let~$\omega$ be a quasi-free state described by the density operator~$W$
(as in Lemma~\ref{lemmaWdef}). We introduce the density operator of the subsystem by
\[ W_A := \tr_{\Fock_B} (W) \::\: \Fock_A \rightarrow \Fock_A \:. \]

\begin{Thm} \label{thmentanglement} The von Neumann entropic difference
\[ S_A(W) := -\tr_\Fock \big(W_A \log W_A) - \tr_\Fock \big( W \log W \big) \]
can be expressed in terms of the reduced one-particle density operator~$D$ by
\[ S_A(W) := \tr \Big( 
\eta \big( \pi_{\H_A} \:D\: \pi_{\H_A} \big) -\pi_{\H_A} \,\eta(D)\,\pi_{\H_A}
\Big) \:, \]
where~$\eta$ is again the von Neumann entropy function~\eqref{etadef},
and~$\pi_{\H_A} : \H \rightarrow \H_A$ is the orthogonal projection operator.
\end{Thm} \noindent
This theorem follows immediately by combining Theorem~\ref{thmreduced} with the following lemma.
\begin{Lemma} The state~$W_A$ is again quasi-free. It is described by the one-particle density operator
\[ D_A := \pi_{\H_A}\, D \,\pi_{\H_A} \; \in \Lin(\H_A) \:, \]
\end{Lemma}
\Proof Let~$\Psi_1, \ldots, \Psi_k$ be any field operator on~$\F_A$. Then
\[ \omega_A \big(\Psi_1 \cdots \Psi_k \big)
= \tr_{\Fock_A} \big( W_A\, \Psi_1 \cdots \Psi_k \big)
= \tr_{\Fock} \big( W\, \tilde{\Psi}_1 \cdots \tilde{\Psi}_k \big) \:, \]
where the tilde denotes the extension to the Fock space~$\F$ by
\[ \tilde{\Psi} \big( \psi_A \otimes \psi_B \big) := \big( \Psi \, \psi_A \big) \otimes \psi_B \:. \]
This shows that the state~$\omega_A$ is again quasi-free. The reduced one-particle density
can be computed using~\eqref{omega2def} and~\eqref{Ddef} for any~$\psi, \phi \in \H_A \subset \H$ by
\[ \langle \psi \,|\, D \phi\rangle_{\H_A}
= \omega_{A,2} (\overline{\psi}, \phi) = \omega_2(\overline{\psi}, \phi) 
= \langle \psi \,|\, D \phi \rangle_{\H} = \langle \psi \,|\, \pi_{\H_A}\, D\, \pi_{\H_A} \phi\rangle_{\H_A} \:. \]
This concludes the proof.
\QED

We finally generalize the previous results to the von Neumann {\em{relative}} entropy. 
For general quasi-free states formulas have been given in terms of the linear complex structure~\cite{hackl-bianchi}. Below, we prove derive in detail how to obtain the relative entropy of particle-number preserving quasi-free states from the reduced one-particle density operator.
\begin{Thm} \label{thmrelreduced}
Let~$W$ and~$W_0$ be the statistical operators of two quasi-free fermionic states with
corresponding reduced one-particle density operators~$D$ and~$D_0$, respectively.
Then the von Neumann relative entropy of~$W$,  in terms of~$D$ and~$D_0$, is given by
\begin{align*}
&S^\rel(W,W_0) : = -\tr_\Fock \big(W \:(\log W-\log W_{0}) \big) \\
&= -\tr_{\H} \Big(  D \big( \log D - \log D_0 \big) + (\1-D)\, \big( \log ( \1 - D)
- \log( \1 - D_0 ) \big)\Big) \:.
\end{align*}
\end{Thm}
\Proof As shown in Theorem~\ref{thmreduced},
\[ \tr_\Fock \big(W \log W) = \tr_{\H} \big( D \log D + (\1-D)\, \log(\1-D) \big) \:. \]

Using again~\eqref{statedef}, we express~$W$ and~$W_0$ in terms of~$D$ and~$D_0$, respectively.
Clearly, since~$D$ and~$D_0$ in general do not commute, we can only diagonalize one
of these matrices, i.e.,
\begin{align*}
W &= \det(\1-D) \:\exp \bigg( -\sum_{n=1}^N s_n\, \Psi_n^\dagger \Psi_n \bigg) \\
W_0 &= \det(\1-D_0) \:\exp \bigg( -\sum_{k,l=1}^N \log \big(D_0^{-1}-\1\big)^k_l \:\Psi_k^\dagger \Psi_l \bigg) \:.
\end{align*}
Thus
\begin{align*}
&\tr_\Fock \big(W \log W_0) = \log \det(\1-D_0) \\
&\qquad\: - \det(\1-D) \: \tr_\Fock \bigg( 
\exp \bigg( -\sum_{n=1}^N s_n\, \Psi_n^\dagger \Psi_n \bigg)
\sum_{k,l=1}^N \log \big(D_0^{-1}-\1\big)^k_l \:\Psi_k^\dagger \Psi_l \bigg) \:.
\end{align*}
Computing the trace with help of the Wick rules, we get zero unless~$k=l$.
Thus
\begin{align*}
&\tr_\Fock \big(W \log W_0) = \log \det(\1-D_0) \\
&\qquad\: - \det(\1-D) \sum_{k=1}^N \log \big(D_0^{-1}-\1\big)^k_k \:\tr_\Fock \Big\{ \Psi_k^\dagger \Psi_k\: e^{-\sum_{n=1}^N s_n\, \Psi_n^\dagger \Psi_n} \Big\} \:.
\end{align*}
Computing the trace again with the help of the formula~\eqref{focktrace},
we obtain
\begin{align*}
&\tr_\Fock \big(W \log W_0) \\
&= \log \det(\1-D_0) 
- \det(\1-D) \sum_{k=1}^N \log \big(D_0^{-1}-\1\big)^k_k \: e^{-s_k} \prod_{n \neq k} \frac{1}{1-d_n} \\
&= \log \det(\1-D_0) - \sum_{k=1}^N \log \big(D_0^{-1}-\1\big)^k_k \: (1-d_k)\: e^{-s_k} \\
&= \log \det(\1-D_0) - \sum_{k=1}^N \log \big(D_0^{-1}-\1\big)^k_k \: d_k \\
&= \tr_{\H} \Big( \log (\1-D_0) - D \log \big(D_0^{-1}-\1\big) \Big) \\
&= \tr_{\H} \Big( \log (\1-D_0) - D \log \big(D_0^{-1} \big( \1 - D_0 \big) \big) \Big) \\
&= \tr_{\H} \Big( \log (\1-D_0) - D \log D_0^{-1} 
- D \log \big( \1 - D_0 \big)\Big) \\
&= \tr_{\H} \Big( \log (\1-D_0) + D \log D_0
- D \log \big( \1 - D_0 \big)\Big) \\
&= \tr_{\H} \Big(  D \log D_0
+ (\1-D)\, \log \big( \1 - D_0 \big)\Big) \:.
\end{align*}
Here we used that for {\em{commuting}} matrices~$A$ and~$B$,
\[ \log(AB) = \log(A) + \log(B) \qquad \text{and} \qquad \log(A^{-1}) = -\log A \:. \]
Combining these formula gives the result.
\QED

Hence, we can write the von Neumann relative entropy of~$W$ as 
\[ S_\rel(W,W_0)=\tr_\Fock \big(W (\log W-\log W_{0})) \:. \]
Since~$W$ and~$W_0$ are density operators and thus self-adjoint, they posses spectral decompositions. Using Klein's inequality, a direct calculation shows that this quantity is non-negative and only zero if~$W=W_{0}$; see
also~\cite{ohya-petz}. \\[-0.5em]

\Thanks{{{\em{Acknowledgments:}} We would like to thank Alexander V.\ Sobolev for helpful discussions.
M.L.\ gratefully acknowledges support by the Studienstiftung des deutschen Volkes and the Marianne-Plehn-Programm. We would like to thank the ``Universit\"atsstiftung Hans Vielberth'' for support.
R.H.J.\ gratefully acknowledges support by the Wenner-Gren Foundations and, in part, by the Wallenberg Initiative on Networks and Quantum Information (WINQ).
Nordita is supported in part by NordForsk.
S.M.\ was partially supported by the MIUR Excellence Department Project 2023-2027 awarded to the Department of Mathematics of the University of Genoa. \\[1em]

\providecommand{\bysame}{\leavevmode\hbox to3em{\hrulefill}\thinspace}
\providecommand{\MR}{\relax\ifhmode\unskip\space\fi MR }
\providecommand{\MRhref}[2]{%
  \href{http://www.ams.org/mathscinet-getitem?mr=#1}{#2}
}
\providecommand{\href}[2]{#2}

\end{document}